\documentclass[format=acmsmall, review=false, screen=true]{acmart}
\renewcommand\footnotetextcopyrightpermission[1]{} 
\usepackage{booktabs} 
\usepackage{multirow}
\usepackage{subfigure}
\usepackage{amsmath}
\usepackage{amssymb}
\usepackage{epstopdf}
\usepackage{mathrsfs}
\usepackage{color}
\usepackage{colortbl}
\usepackage{lineno}
\DeclareMathOperator*{\argmax}{arg\,max}

\usepackage[ruled]{algorithm2e} 

\SetAlFnt{\small}
\SetAlCapFnt{\small}
\SetAlCapNameFnt{\small}
\SetAlCapHSkip{0pt}
\IncMargin{-\parindent}

\setcopyright{acmcopyright}
\setcopyright{acmlicensed}
\setcopyright{rightsretained}
\setcopyright{usgov}
\setcopyright{usgovmixed}
\setcopyright{cagov}
\setcopyright{cagovmixed}

\acmDOI{0000001.0000001}

\begin{document}
\title{CNNs based Viewpoint Estimation for Volume Visualization}

\author{Neng Shi}
\affiliation{%
  \institution{State Key Lab of CAD\&CG, Zhejiang University}
  \streetaddress{866 Yuhangtang Rd}
  \city{Hangzhou}
  \state{Zhejiang}
  \postcode{310058}
  \country{China}}
\email{shineng@zju.edu.cn}

\author{Yubo Tao}
\authornote{This is the corresponding author}
\affiliation{%
  \institution{State Key Lab of CAD\&CG, Zhejiang University}
  \streetaddress{866 Yuhangtang Rd}
  \city{Hangzhou}
  \state{Zhejiang}
  \postcode{310058}
  \country{China}}
\email{taoyubo@cad.zju.edu.cn}

\begin{abstract}
Viewpoint estimation from 2D rendered images is helpful in understanding how users select viewpoints for volume visualization and guiding users to select better viewpoints based on previous visualizations. In this paper, we propose a viewpoint estimation method based on Convolutional Neural Networks (CNNs) for volume visualization. We first design an overfit-resistant image rendering pipeline to generate the training images with accurate viewpoint annotations, and then train a category-specific viewpoint classification network to estimate the viewpoint for the given rendered image. Our method can achieve good performance on images rendered with different transfer functions and rendering parameters in several categories. We apply our model to recover the viewpoints of the rendered images in publications, and show how experts look at volumes. We also introduce a CNN feature-based image similarity measure for similarity voting based viewpoint selection, which can suggest semantically meaningful optimal viewpoints for different volumes and transfer functions.%
\end{abstract}

%
%


%
%


\maketitle

\renewcommand{\shortauthors}{N. Shi and Y. Tao}

\section{Introduction}
Viewpoint is one of the important rendering parameters in volume visualization,
and it is intentionally selected by users to convey important features clearly
and to meet their aesthetic preferences. Poorly chosen viewpoints can lead to an
imprecise and misleading analysis of volumetrical features; however, it is not always
easy for the general users to choose a good viewpoint from scratch due to the high
degree of freedom. Thus, many automatic viewpoint selection methods, such as surface area entropy~\cite{Takahashi:2005:FAL}, voxel
entropy~\cite{Bordoloi:2005:VSV}, opacity entropy~\cite{Ji:2006:DVS}, and
gradient/normal variation~\cite{Zheng:2011:iView}, have been
proposed to suggest optimal viewpoints to serve as a starting point of volume
exploration. However, relatively little
research has considered the viewpoint estimation problem from a rendered image in volume visualization.
Viewpoint estimation can help us to understand how experts select viewpoints
for volume visualization and guide users to select better viewpoints based on
previously rendered images. It is also the first step in recovering a visual encoding
specification from a rendered image, such as the transfer functions~\cite{Raji:2017:PGE}.

The viewpoint estimation problem can be transformed into a learning problem~\cite{Kendall:2015:PCN, Massa:2016:CMC},
and there are two main challenges in constructing robust models. The first
challenge is the lack of diverse training images with accurate viewpoint annotations.
Although there are many rendered images in published papers on volume visualization,
their viewpoints are unknown, and it is time-consuming and less accurate to
annotate these images manually. In contrast to one feature in a 3D model, there are
many features in a volume which are classified by transfer functions, and each image
may contain only one or some of the features, such as the skin, bone and tooth in the head volume. In addition, other rendering parameters,
such as projection types, also potentially affect the rendered results and
their viewpoint estimation. Therefore, the training images need to be sufficiently diverse
to include different features and rendering parameters.
The second challenge is to design powerful features specially tailored for viewpoint
estimation. SIFT (Scale-Invariant Feature Transform) and HOG (Histogram of Oriented
Gradients) are the two most commonly used features, and they have been used for measurement of image
similarity in the image-based viewpoint selection model~\cite{Tao:2016:SVV}.
However, they are designed primarily for image classification and object detection.
Recently, Convolutional Neural Networks (CNNs) have been shown to automatically learn better features
via task-specific supervision, i.e., the lower layers mostly detect
low-level features, such as corners, color patches and stripes, while the higher layers
aggregate these low-level features into high-level task-related features, such as cats and
automobiles for image classification. CNNs have been used to determine
up front orientations and detect salient views of 3D models~\cite{Kim:2017:CSS}. Therefore,
CNNs are an attractive choice for extracting specific features for viewpoint estimation in
volume visualization.

In this paper, we propose a viewpoint estimation method based on CNNs. Since CNN training
requires a huge amount of viewpoint-annotated images, we design an overfit-resistant image rendering pipeline, inspired by the ``Render for CNN'' idea~\cite{Su:2015:RCV},  to
generate the training dataset. Many volumes are available online in large public
volume collections, and they can be classified into several categories, such as the head
and tree. Given a category, we take into account different features and rendering
parameters to generate diverse training images with accurate viewpoint annotations.
After that, we train a category-specific viewpoint classification network to estimate the viewpoint
of a rendered image in this category. Our method can achieve good performance on images
rendered with different transfer functions and rendering parameters.

We present two applications of our viewpoint estimation method. The first application
investigates how visualization experts select viewpoints for volume visualization.
For the collected images in the volume visualization literature, we estimate the
viewpoints of these images using our viewpoint estimation network, and analyze
how visualization experts look at the volume in different categories. Inspired by
the image-based viewpoint selection model, our second application
suggests an optimal viewpoint with a clear semantic meaning for general users. We
introduce a CNN-feature based image similarity measure and apply the measure to the
similarity voting based viewpoint selection. Thus, our method can suggest different
viewpoints for different volumes and transfer functions based on the similarity
between collected images in the volume visualization literature and rendered images.

In summary, our main contributions are as follows:
\begin{itemize}
\item {
A CNN based viewpoint estimation method for volume visualization. We propose an
overfit-resistant image rendering pipeline to generate the training dataset considering
different transfer functions and rendering parameters which are as diverse as possible, and we design
a geometric structure-aware loss function customized for viewpoint estimation.
}
\item {
Two applications of our viewpoint estimation method: an analysis of the viewpoint preferences
 of visualization experts, and viewpoint selection based on a CNN-feature
based image similarity measure.
}
\end{itemize}

\section{Related Work}

The viewpoint estimation problem can be considered to be an ``inverse problem''
of data visualization, i.e., given a visualization, can we recover the underlying
visual encoding and even data values? This is useful for automated analysis,
indexing and redesign of previous visualizations. This paper focuses on recovering
the viewpoint from a rendered image. Thus, we review the related work on viewpoint
selection, reverse engineering of visualizations, and pose estimation.

\subsection{Viewpoint Selection}

Viewpoint selection has been widely investigated in computer graphics and
visualization, and is a forward problem of viewpoint estimation.
Computer-graphics psychophysics provided the ``canonical views''~\cite{Blanz:1999:WOA},
a small number of user-preferred viewpoints with the attributes of goodness for
recognition, familiarity, functionality, and aesthetic criteria. Thus, the optimal
viewpoint often presents the most information about features of interest. V\'{a}zquez
et al.~\cite{Vazquez:2001:VSU} first applied information theory to search for the optimal
viewpoint. Polonsky et al.~\cite{Polonsky:2005:WTC} proposed three principles (view-independent,
view-dependent, and semantic meaning) to classify and compare view descriptors for
3D models. Wang et al.~\cite{Cremanns:1994:CCP}  proposed a search strategy for viewpoint selection by identifying the regions that are very likely to contain best views, referred to as canonical regions, attaining greater search speed and reducing the number of views required.  Wu et al.~\cite{journals/corr/WuSKTX14} proposed to represent a geometric 3D shape as a probability distribution of binary variables on a 3D voxel grid, and this model is able to predict the next-best-view for an object. 


In volume visualization, Takahashi et al.~\cite{Takahashi:2005:FAL} decomposed the volume into
features, calculated the optimal viewpoint for each feature using the surface area
entropy, and combined these optimal viewpoints to suggest the optimal viewpoint for
all features. Bordoloi and Shen~\cite{Bordoloi:2005:VSV} proposed the voxel entropy
to identify representative viewpoints, and Ji and Shen~\cite{Ji:2006:DVS} further
presented image-based metrics, including opacity entropy, color entropy and curvature
information. Ruiz et al.~\cite{Ruiz:2010:VIC} introduced the voxel mutual information
to measure the informativeness of the viewpoint. A viewpoint suggestion framework
presented by Zheng et al.~\cite{Zheng:2011:iView} first clusters features based on
gradient/normal variation in the high-dimensional space, and iteratively suggests
promising viewpoints during data exploration. 



A growing body of work focuses on the data-driven or learning based viewpoint selection
methods. Vieria et al.~\cite{Vieira:2009:LGV} presented intelligent design galleries
to learn a classifier based on a large set of view descriptors from the user interaction
on viewpoints. Secord et al.~\cite{Secord:2011:PMV} collected the relative goodness of
viewpoints based on human preferences through a user study, and trained a linear model
based on these collected data for viewpoint selection. A web-image voting method proposed
by Liu et al.~\cite{Liu:2012:WDB} allows each web-image to vote its most similar
viewpoints based on the image similarity considering the area, silhouette and saliency attributes, and
it performs better than previous view descriptors for 3D models. Since visualization
experts generally provide more representative viewpoints for volumes, Tao et
al.~\cite{Tao:2016:SVV} utilized rendered images in published papers on volume
visualization to learn how visualization experts choose representative viewpoints for
volumes with similar features. The viewpoint voting is based on the image similarity
with SIFT and HOG between the collected image in published papers and the rendered image under the same
viewpoint. Our viewpoint selection method is also based on the similarity voting,
but the image similarity is evaluated based on learned features from CNNs, not
manually designed features.

\subsection{Reverse Engineering of Visualizations}

Most research on reverse engineering of visualizations focus on static chart images, such as line charts,
pie charts, bar charts and heatmaps, and many methods have been proposed to
interactively or automatically extract data values and encoding specifications
for visualization interpretation and redesign.

ReVision~\cite{Savva:2011:ReVision} automatically identifies the chart type of
bitmap images, infers the data by extracting the graphical marks, and redesigns
visualizations to improve graphical perception. Harper and Agrawala~\cite{Harper:2014:DRD}
presented a deconstruction tool to extract the data in a D3 visualization and allow users to
restyle existing D3 visualizations. FigureSeer~\cite{Siegel:2016:FigureSeer}
applies a graph-based reasoning approach based on a CNN-based similarity metric
to extract data and its associated legend entities to parse figures in
research papers. iVoLVER~\cite{Mendez:2016:IVoLVER} enables flexible data
acquisition from bitmap charts and interactive animated visualization
reconstruction. Jung et al.~\cite{Jung:2017:ChartSense} introduced ChartSense
to determine the chart type using a deep learning based classifier and to
semi-automatically extract data from the chart image. Instead of data values,
Poco and Heer~\cite{Poco:2017:REV} automatically recovered visual encodings
from a chart image based on inferred text elements. They further contributed
a method to extract color mapping from a bitmap image semi-automatically,
and presented automatic recoloring and interactive overlays to improve
perceptual effectiveness of visualizations~\cite{Poco:2018:ERC}.

Besides chart images, there are several learning-based methods to recover
the viewpoint and transfer function from a rendered image of 3D models and
volumes. Liu et al.~\cite{Liu:2016:UOS} described a data-driven method for
3D model upright orientation estimation using a 3D CNN. Similarly, Kim et
al.~\cite{Kim:2017:CSS} applied one CNN on 3D voxel data to generate a CNN
shape feature for the upright orientation determination, and the other CNN
to encode category-specific information learned from a large number of 2D
images on the web for the salient viewpoint detection. Given a target image,
Raji et al.~\cite{Raji:2017:PGE} combined CNN and evolutionary optimization
to iteratively refine a transfer function to match the visual features in
the rendered image of a similar volume dataset to the one in the target image.
Their CNN, is used to compare the similarity between the rendered and target image, not trained specially for the transfer function optimization task. In this paper,
our CNN is an end-to-end training for viewpoint estimation. 


\subsection{Pose Estimation}

Pose estimation is an active branch of research in computer 
vision for object detection and scene understanding. For example, the indoor mapping problem is based on estimating the pose of the sensor of each k-th frame and building a map of the environment with the estimated camera pose of each frame~\cite{Figueroa:2015:CAT}. Recently, most methods have been based on CNNs,
and these methods can be divided into two categories: keypoint-based method and direct estimation method.

The keypoint-based method usually predicts 2D keypoints from an image, and recovers
the 3D pose from these keypoints by solving a perspective-n-point problem. These 2D
keypoints can be semantic keypoints defined on 3D object models~\cite{pavlakos2017object3d, wu2016single}.
Given an image, the CNN trained on semantic keypoints is used to predict a probabilistic map of
2D keypoints and recover the 3D pose by comparsion with pre-defined object models. Instead of
semantic keypoints, 2D keypoints can be eight corners of the 3D bounding box encapsulating the 
object~\cite{DBLP:journals/corr/abs-1803-11493, Rad2017BB8AS}. The CNN is trained by comparing the 
predicted 2D keypoint locations with the projections of 3D corners of the bounding box on the 
image under the ground-truth pose annotations.

The direct estimation method predicts the 3D pose from an image without intermediate keypoints, and mostly uses the Euler
angle representation of rotation matrices to estimate the azimuth, elevation and camera-tilt angles
separately. The pose estimation problem can be solved by directly regressing the angle with a
Euclidean loss~\cite{Wang2016Regression}, or through transformation into a classification problem by dividing 
the angle into non-overlapping bins~\cite{tulsiani2015viewpoints, Su:2015:RCV, elhoseiny2016comparative}. 
Massa et al.~\cite{Massa:2016:CMC} experimented with multiple loss functions in CNNs based on regression and 
classification, and concluded that the loss function based on classification outperforms the one based on
regression by a considerable margin. They further proposed a joint object detection and viewpoint 
estimation method for diverse classes in the Pascal3D+ dataset~\cite{xiang2014beyond}. Besides the Euler angle
representation, PoseCNN~\cite{Xiang:2017:PoseCNN} employs the quaternion representations of 3D rotations, 
introduces a new loss function for the 3D rotation regression problem to handle symmetric objects, and 
estimates 6D object pose in cluttered scenes.  Mahendran et al.~\cite{Mahendran2018AMC} proposed an
axis-angle representation in a mixed classification  regression framework. This framework can accommodate different  architectures and loss functions to generate multiple classification-regression models, and it achieves good performance on the Pascal3D+ dataset.



Similarly, our method is also based on CNNs. However, pose estimation in
computer vision mostly focuses on analyzing the localization of the object in the real scene
for mobile robotics, navigation and augmented reality. Our objective is to estimate the
viewpoint of a rendered image to recover the rendering parameters. 
Technically, there are two differences between pose estimation and viewpoint estimation. The first is the camera's intrincic parameters.  Pose estimation is generally based on the assumption that the camera's intrinsic parameters are known. For example, the ground-truth poses in the Pascal3D+ dataset are computed from 2D-3D correspondences assuming the same intrinsic parameters for all images. However, this paper attempts to estimate the viewpoint of a rendered image under different intrinsic parameters of the camera, especially different projection types (parallel projection and perspective projection). Thus, the keypoint-based method is not suitable for our problem, since it is difficult to solve the perspective-n-point problem with the unknown intrinsic parameters of the camera. 
The second is that pose estimation aims to predict the 3D rotation between the object and the camera. Under the Euler angle representation, the 3D pose includes azimuth, elevation and camera-tilt angle. In this paper, we only concern about the camera's viewpoint, including only azimuth and elevation under the Euler angle representation, and the camera-tilt angle is less interesting in our viewpoint estimation for volume visualization.  As a result, when we apply the direct estimation method in pose estimation to our viewpoint estiamtion, we need to revise the angle representation at first. 


\section{Viewpoint Estimation}

Given an input rendered image, our goal is to estimate the viewpoint. We assume that all
viewpoints are on the viewing sphere~\cite{Ji:2006:DVS}, the center of which is located
at the volume center. We can parameterize the viewpoint as a tuple ($\theta$, $\phi$) of
camera parameters, where $\theta$ is the azimuth (longitude) angle and $\phi$ is the
elevation (latitude) angle. The viewpoint estimation problem can be transformed to a
regression problem. However, the regression model only returns predicted camera parameters, 
and may not capture the underlying viewpoint ambiguity, such as similar rendered images of
nearby viewpoints for some volumes and symmetrical viewpoints for semi-transparent volumes.
On the other hand, we can divide the continuous camera parameter domain into intervals
and transform it into a classification problem. Thus, we can obtain the probabilities of
each interval after the classification. Experiments also show that the classification performance is better than the regression performance for viewpoint estimation~\cite{Massa:2016:CMC}.

Previous classification methods for viewpoint estimation~\cite{Su:2015:RCV, tulsiani2015viewpoints, elhoseiny2016comparative}
divide the azimuth and elevation domain independently, and the loss is simply the sum of the azimuth and elevation misclassification.
However, the azimuth and elevation are not uniform units of measure, similar to the longitude and latitude of the earth, and 
they can not be directly used to evaluate the distance between the predicated and ground-truth viewpoint. In order to overcome this problem, 
this paper explicitly divides the viewing sphere into $N$ uniform regions and assigns a viewpoint label for each region. 
Thus, the viewpoint estimation problem can be formalized as classifying the rendered image into viewpoint labels with probabilities, 
and the loss is evaluated by the geodesic distance between the predicted and ground-truth viewpoint. 



\begin{figure*}[tbp]
  \center
  \includegraphics[width=1\linewidth]{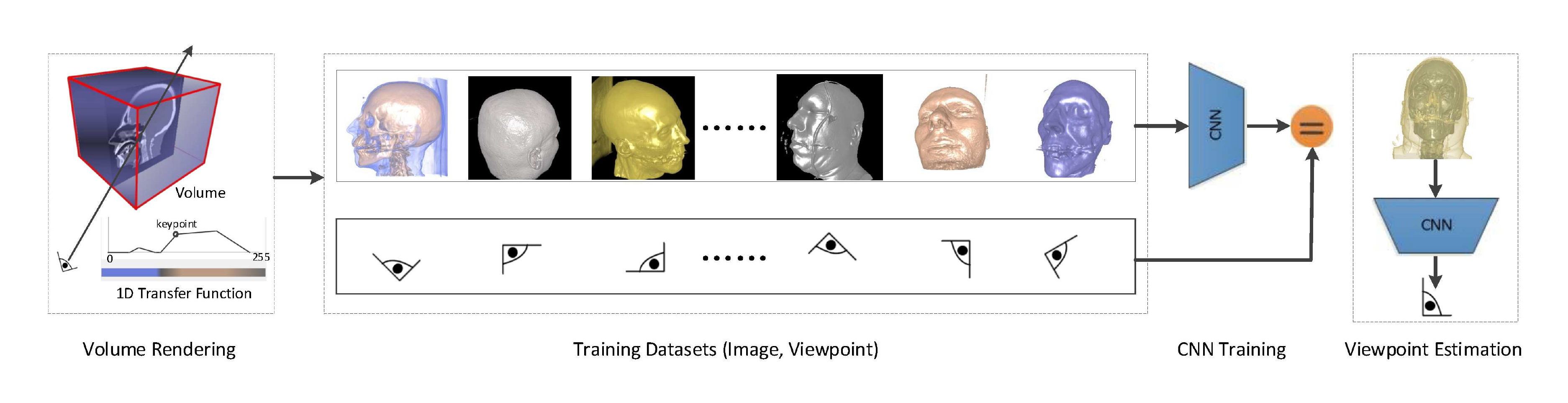}
  \caption{The viewpoint estimation learning pipeline. For each category (heads in
           this example), we apply direct volume rendering with the volumes, transfer functions,
           and other rendering parameters as the input  to generate the training datsets
           (rendered images with annotated viewpoints). The CNN training process takes
           rendered images as input, to estimate the viewpoint, and the parameters of the
           CNN are optimized by minimizing the difference between the estimated viewpoint and the annotated
           viewpoint. Finally, the trained CNN can be used to estimate the viewpoint of an
           image, rendered with a different volume in this category, different transfer
           functions and rendering parameters.}
  \label{fig:pipeline}
\end{figure*}

We apply CNNs to the viewpoint classification
problem due to their high learning and classification capacities. A large number of
viewpoint-annotated images of high variation are required to avoid overfitting of deep
CNNs. Since there is no training dataset available for estimating the viewpoints of rendered 
images, we generate the training dataset through ``Render for CNN'' ~\cite{Su:2015:RCV} approach on the 
volumes in large public collections. The rendering process is more complex for volumes than 
 for 3D models, since both the data classification and the rendering parameters have a strong influence 
on the rendered image. As shown in Fig.~\ref{fig:pipeline}, we first classify volumes available
online into several categories, and design an overfit-resistant image rendering pipeline
to generate the training dataset. This pipeline should consider both the different data classification and the different rendering parameters. After obtaining these rendered images with viewpoint
labels, we train a category-specific viewpoint classification network for each category.
With the trained CNN, we can estimate the viewpoint of a new rendered image of volumes in these categories.

\subsection{Training Image Generation}
\label{sec:generation}

The viewpoints are sampled uniformly on the viewing sphere by Hierarchical Equal Area isoLatitude Pixelization (HEALPix)~\cite{Gorski:2005:HAF}, which can effectively discretize a sphere, and these viewpoints are our viewpoint labels. There are several volumes or one volume in each category in the training dataset depending on the volumes available online. For example, the head category has three different volumes, and the engine category has only one volume. We design an overfit-resistant image rendering pipeline to generate rendered images with viewpoint annotations as the training dataset for each category.

For each volume in the category, we render as many images as possible for each viewpoint label according to different data classification and rendering parameters. For each rendered image, the rendered viewpoint is randomly shifted within the region of the viewpoint label to avoid overfitting in the training process.

\textbf{Data classification}. The transfer function classifies the features in the volume~\cite{Khan:2018:NIA}. Different opacities are specified to highlight features of interest and remove unrelated features. Different colors are used to label different features. Thus, the transfer function has a strong influence on the rendered image and its viewpoint classification. During training, various transfer functions are required to generate rendered images with as many different features as possible.

We manually design different opacity transfer functions for each volume to classify different features. For example, the head volume generally has the skin, skull, and tooth, and the opacity transfer functions are designed to show only one feature semi-transparently or opaquely, or some of the features with semi-transparent outer features and semi-transparent or opaque inner features. Opacity transfer functions are not randomly generated in our training dataset, since random opacity transfer functions may easily miss important features completely, and these rendered images may lack features and reduce the performance of the viewpoint estimation. Our model is expected to estimate the viewpoint from the rendered image generated from a manually designed opacity transfer function, such as the rendered image in the visualization paper, instead of a random opacity transfer function. In order to improve the generalization, we still add a small random disturbance to the designed opacity transfer function. For each rendered image, we randomly adjust the opacity slightly for each feature independently, such as moving the keypoint of the 1D transfer function (Fig.~\ref{fig:pipeline}) left or right by the distance $d \sim \mathcal{N}(0,1)$. For the color transfer function, the color of each feature is randomly sampled for each rendered image, since users may choose different colors for features during data classification. The color is biased towards a high contrast with the background color to mimic a user's intent on emphasizing features. These random color transfer functions would improve the generation of viewpoint estimation. 

\textbf{Rendering parameters}. Besides the viewpoint, there are many other rendering parameters: the camera-tilt angles, scales, projection types, background color, and so on. Since CNNs are not rotation invariant, i.e., if the whole image is rotated then the CNNs' performance suffers, we need to deal with rotation invariance through data augmentation, i.e., the effect of the camera-tilt angle.
We randomly rotate \emph{the camera-tilt angle} for each rendered image. There are generally two projection types, parallel projection and perspective projection. Thus, for each rendered image, \emph{the projection type} is randomly selected from the two projection types. The background color also affects features due to alpha blending in direct volume rendering. The most common background colors are black and white, and we randomly choose black or white as \emph{the background color} for each rendered image. It is worth noting that the color transfer functions in different backgrounds are slightly different in order to distinguish the features from the background. Although CNNs are relatively invariant to scaling, we further reduce the influence of \emph{the scale} by rendering volumes with the scale uniformly sampled from 1 to 1.8.

For \emph{the lighting condition}, three lighting modes are used. The first is the environment light only. The second is the environment light and one headlight located at the same position of the camera, and the lighting intensity is uniformly sampled from 0.7 to 1. The third one includes environment light, one headlight, and one scene light. The position of the scene light is uniformly sampled on a sphere with the radius uniformly sampled from 3 to 5 times the radius of the viewing sphere, and the lighting intensities of the headlight and scene light are uniformly sampled from 0.35 to 0.5. The coefficients of the Phong reflection model are also randomly sampled. The ambient reflection coefficient is fixed at 1, the diffuse reflection coefficient is uniformly sampled from 0.25 to 0.75, the specular reflection coefficient is uniformly sampled from 0.5 to 1, and the shininess coefficient is uniformly sampled from 20 to 100. Other rendering parameters, such as gradually changed background colors, can be included in our image rendering pipeline to further improve the generalization of viewpoint estimation.

\subsection{Network Architecture and Loss Function}
\label{sec:network}

In this section, we introduce the network architecture and different loss functions for viewpoint estimation problem, standard cross-engropy loss function and geometric structure-aware loss function respectively. 


\subsubsection{Network Architecture} 

A CNN is a type of artificial neural network, and has become a hotspot in computer vision and natural language processing due to its satisfactory learning capacity. For instance, it has been applied to question answering (QA) systems~\cite{Wang:2018:CAC} and image recognition~\cite{Ou:2017:AIV}. It is a multilayer perceptron specifically designed to recognize 2D shapes, and this network structure is invariant to translation, scaling, or other forms of deformation~\cite{Goodfellow:2016:DP}.

\begin{figure*}
	\center			
    \subfigure[] {\label{fig:newwork}  \includegraphics[width=0.74\linewidth]{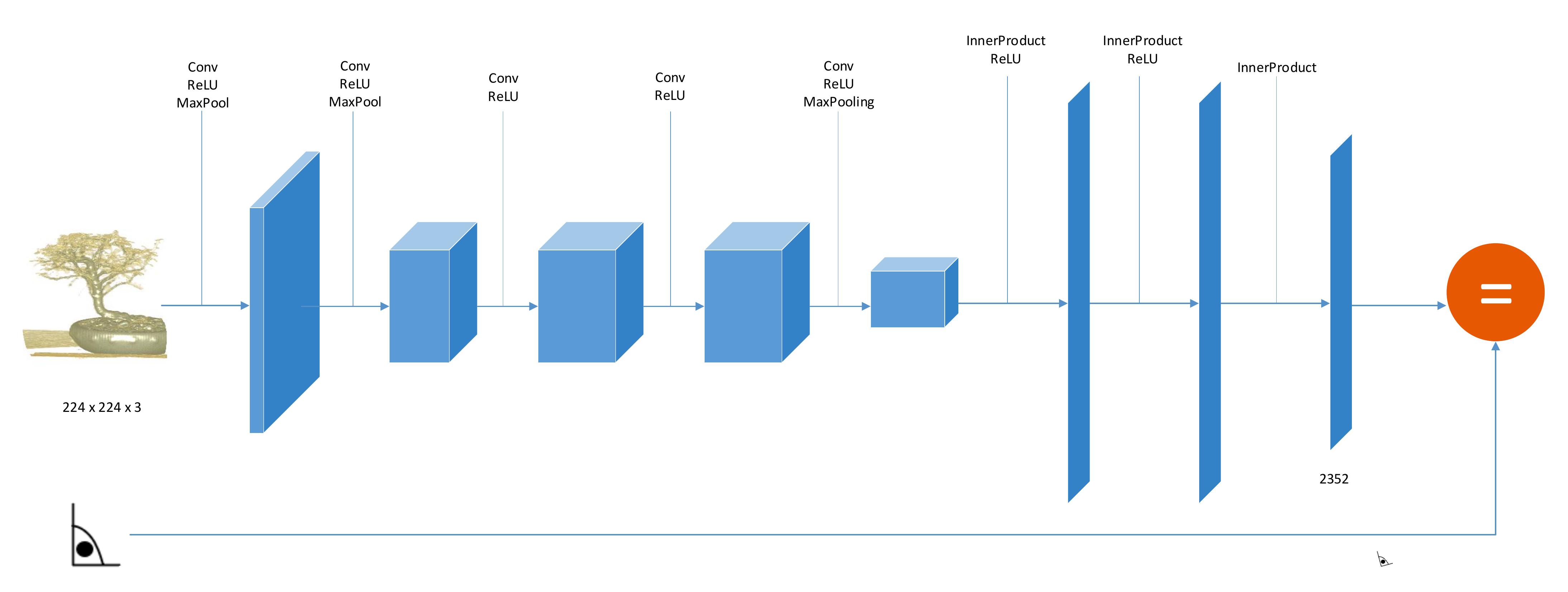}}
    \subfigure[] {\label{fig:neighbor} \includegraphics[width=0.24\linewidth]{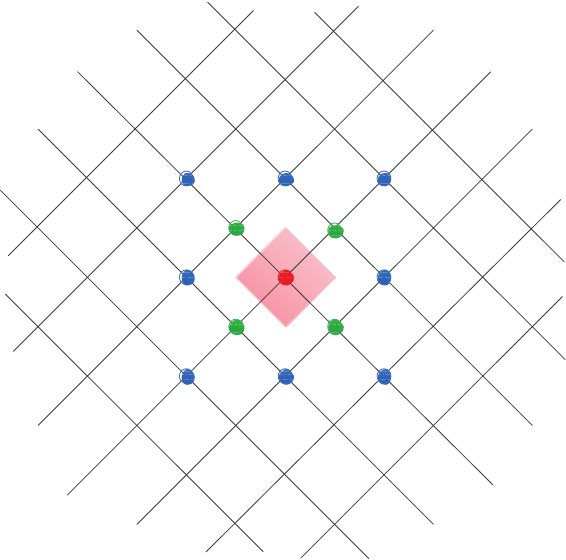}}
    \caption{(a) The structure of our viewpoint estimation network based on AlexNet. 
    		(b) The illustration of neighbor viewpoints in $V(v_s, n)$  
			for the geometric structure-aware loss function. The
			red viewpoint is the ground-truth viewpoint $v_s$, and its first-order neighbor viewpoints are the
			four green viewpoints ($n=1$).}
	\label{fig:pl}
\end{figure*}

Recently, many well-designed networks have been proposed, such as AlexNet~\cite{Krizhevsky:2012:ICW}, VGGNet~\cite{simonyan2014very}, GoogLeNet~\cite{Szegedy:2014:GDC} and ResNet~\cite{He:2016:DRL}, and they all achieve good performance in image classification. We first experiemnt with AlexNet, and then mainly choose the 19-layer VGGNet to implement the viewpoint classification task.
The structure of our network based on AlexNet is shown in Fig.~\ref{fig:newwork}. The network ends with an N-way fully-connected layer with softmax, according to the assigned viewpoint label. 





\subsubsection{Geometric Structure-Aware Loss Function}

The loss function in the last layer is very important for viewpoint estimation. The widely used loss
function, Softmax loss, employs the output probability as the predicted probability and computes the cross entropy loss based on the ground-truth value. During training, minimizing the cross entropy is equivalent to maximizing the log-likelihood of the ground-truth label. The disadvantage of the Softmax loss function is that it learns to predict the viewpoints without explicitly considering the continuity between neighbor viewpoints. It is obvious that two neighbor viewpoints have great deal in common, and the geometric information may be particularly important for viewpoint estimation. One solution to the problem is to design a geometric structure-aware loss function customized for viewpoint estimation.

The geometric structure aware loss function is modified from the Softmax loss function by adding geometric constraints:
\begin{equation}
L_{vp} (\{s\}) = -\sum\limits_{\{s\}} \sum\limits_{v\in V(v_s,n)} q(v) logP_v(s),
\end{equation}
where $v_s$ is the ground-truth viewpoint for the rendered image $s$. $V(v_s,n)$ is the neighbor viewpoint
set of the ground-truth viewpoint $v_s$, determined by the relative distance bandwidth parameter $n$. Since HEALPix can provide a viewpoint's neighbors, the neighbor viewpoint set contains the viewpoint itself for $n=0$, the viewpoint and its first-order neighbor viewpoints for $n=1$, and so on. For example, Fig.~\ref{fig:neighbor} shows the groud-truth viewpoint and its neighbor viewpoints on the viewing sphere. The light red area is the region
for the red viewpoint, i.e., all viewpoints in the light red area have the same label as the red viewpoint.
The four green viewpoints are the first-order neighbor viewpoints of the red viewpoint.
$P_v(s)$ is the probability for the image $s$ classified to the viewpoint label $v$ based on the Softmax loss function. 

The only difference between the geometric structure-aware loss function and the orginial Softmax loss function is the designed ground-truth distribution $q(v)$:
\begin{equation}
q(v) = e^{-d(v,v_s)},
\end{equation}
where $d:V \times V \mapsto \mathbb{R} $ is the geodesic distance between two viewpoints. 
We substitute an exponential decay weight w.r.t the
viewpoint distance, to explicitly exploit the correlation between neighbor viewpoints.
In our experiment, the relative distance bandwidth parameter is 1 ($n$ = 1), i.e., we consider only
the ground-truth viewpoint and its first-order neighbor viewpoints. 
 In the Softmax loss function, 
$q(v_s) = 1$ and $q(v) = 0$ for $v \neq v_s$. However, in the geometric structure-aware loss function, 
$q(v_s) = 0.87$,  $q(v) = 0.36$ for $v\neq v_s$ and $v \in V(v_s,1)$, otherwise $q(v) = 0$ for $v \notin V(v_s,1)$. We expect the geometric structure-aware loss function with the bandwidth parameter as 1 could improve the prediction accuracy, compared with the original softmax loss function.
\section{Results}

\subsection{Training Datasets}
\label{sec:datasets}

{\fontsize{9pt}{9pt}\selectfont
\begin{table}
\newcommand{\tabincell}[2]{\begin{tabular}{@{}#1@{}}#2\end{tabular}}
\caption{The statistical information for the training dataset in each category: the number of opacity transfer functions, the number of rendered images ({$\times10^5$}), and main features classified by opacity transfer functions.}
\begin{tabular}{|c|c|c|c|}
\hline
Category & Transfer Functions & Images & Main Features \\
\hline
engine & 9 & 4.5 & surface, gear, other inner structure\\
\hline
fish & 6 & 3.0 & bone,  skin, gill\\
\hline
head & 12 & 6.0 & skull, skin, tooth\\
\hline
tooth & 5 & 2.5 &  enamel, dentin,  pulp chamber, cementum  \\
\hline
tree & 7 & 3.5 & trunk, branch, leaf \\
\hline
vessel & 2 & 1.0 &  vessel, aneurism\\
\hline
\end{tabular}
\end{table}
}

We collected available volumes from public volume databases, such as VolVis and The Volume Library,
and extracted six volume categories manually: engine, fish, head, tooth, tree, and vessel. Most of these volumes are widely used in volume visualization research. There are three different volumes in the head and tree categories, and the other categories have only one volume. Since the head and tree categories have more than one volume, the experiment can demonstrate the generalization of our classification model when there are multiple volumes in the category.

In our experiment, the number of viewpoints is $N=2,352$, and we apply the proposed image rendering pipeline to generate training images with viewpoint annotations for each category. As shown in Table 1, we take into consideration main features of each category when we manually design these opacity transfer functions. For example, the tree generally has the trunk, branch and leaf features, and the vessel only has the vessel and aneurism features. Thus, the number of opacity transfer functions is different for each category depending on the number of volumes and features in each volume. For instance, there are seven, four and one opacity transfer functions for the Chapel Hill CT head, the visual male, and the MRI head in the head category, respectively, and there are five opacity transfer functions for different features for the tooth volume in the tooth category. We rendered 50,000 images at 256 $\times$ 256 size for each opacity transfer function, considering different viewpoints, color transfer functions, camera-tilt angles, projection types, and lighting conditions. The number of training images is from one hundred thousand to six hundred thousand, as listed in Table 1.


\subsection{Training Process}

With these rendered images, we train a viewpoint classification network for each category. Because of the
abundant labeled training data in ImageNet, pretrained models on ImageNet would generally have a very powerful generalization ability. The low and middle layers contain a massive number of general visual elements, and we only need to finetune the last several layers based on our training dataset for the viewpoint classification task. Thus, for both AlexNet and the 19-layer VGGNet, convolution layers from conv1 to conv3 are fixed with the pretrained parameters. The remaining convolutional layers and all the fully connected layers except the last one are finetuned during the training process. Only the last fully connected layer is trained from scratch.

The network is implemented in Caffe~\cite{Jia:2014:Caffe}, using an NVIDIA GTX 1080 Ti GPU. For the 19-layer VGGNet, the network is trained by  stochastic gradient descent of about 2 epochs, and the total training time is about 3 days for each category.
During the testing, the viewpoint estimation time of each rendered image is about 0.01 seconds based on the proposed network.

\subsection{Viewpoint Estimation Evaluation}

We first describe how to evaluate the accuracy of viewpoint estimation. The trained CNN generates
a probability for each viewpoint label, and for most cases, the classification probability distribution on the viewing sphere approximately subjects to a bivariate Gaussian distribution. Thus, we can model the distribution through $P_v(s) \sim N(v_\mu,v_\sigma^2)$, where $v_\mu$ equals
\begin{equation}
\argmax_{v \in V}P_v(s).
\end{equation}
The mean $v_\mu$ can be used to evaluate the accuracy of our viewpoint classification network. 
The standard deviation $v_\sigma$ is small, when the CNN is very confident to estimate the viewpoint with a high probability.
For challenging cases, the CNN becomes less confident and the $v_\sigma$ becomes bigger. 
Examples in the head category are shown in the first row of Fig.~\ref{fig:badClassification}. 
The first two are simple cases with a small deviation and the last two are challenging cases with a large deviation.  

We define the accuracy metric based on the geodesic distance between the estimated viewpoint $v_\mu$ and the ground-truth viewpoint $v_s$. Our evaluation metric is a viewpoint accuracy with a ``tolerance". Specifically, we select five tolerances, $2\ ^{\circ}$, $5\ ^{\circ}$,  $8\ ^{\circ}$, $11\ ^{\circ}$ and $15 ^{\circ}$, respectively. In the evaluation, if the geodesic distance between $v_\mu$ and $v_s$ is within the tolerance, we count it as a correct prediction. When the geodesic distance is within $2\ ^{\circ}$, the predicted viewpoint $v_\mu$ exactly matches the ground-truth viewpoint $v_s$.   

In our evaluation, we apply the same image rendering method in Sec.~\ref{sec:generation} to generate the testing dataset. Since we add a random disturbance for each opacity transfer function and randomly sample one color for each feature, the transfer functions in the testing dataset are different from the ones in the training dataset. 
We generate 3,000 images for each opacity transfer function in each category, for example 27,000 ($3,000 \times 9$) images for the engine category.

{\fontsize{9pt}{9pt}\selectfont
\begin{table}
\newcommand{\tabincell}[2]{\begin{tabular}{@{}#1@{}}#2\end{tabular}}
\caption{Classification accuracy comparison of AlexNet (UD+GS, uniform division of the viewing sphere and geometric structure-aware loss function), VGGNet (UD+GS), VGGNet (UD+Softmax, uniform division of the viewing sphere and Softmax loss function), VGGNet (SD+GS, separate division of the azimuth and elevation and geometric structure-aware loss function) and the category-independent network VGGNet.CI (UD+GS) on the six categories under different tolerances.}
\begin{tabular}{|c|c|c|c|c|c|c|}
\hline
Cat. & Angle Tol. & $2^{\circ}$ & $5^{\circ}$ & $8^{\circ}$ & $11^{\circ}$ & $15^{\circ}$  \\
\hline
\multirow{5}*{engine} 
& AlexNet (UD+GS)  & 0.4664 & 0.8194 & 0.9592 & 0.9860 & 0.9931  \\
\cline{2-7}
~ &  VGGNet (UD+GS) & \textbf{0.8450} & \textbf{0.9774} &  \textbf{0.9987} &  \textbf{0.9997} & \textbf{0.9999}  \\
\cline{2-7}
&  VGGNet (UD+Softmax) &  0.6692 &  0.9313 &  0.9942 & 0.9985 & 0.9996  \\
\cline{2-7}
&  VGGNet (SD+GS) & 0.5371 & 0.9412 & 0.9896  & 0.9948 & 0.9957  \\
\cline{2-7}
& VGGNet.CI (UD+GS) & 0.6845 & 0.9279 & 0.9911 & 0.9982 & 0.9995  \\
\hline

\multirow{5}*{fish} 
& AlexNet (UD+GS) & 0.5859 & 0.8315 & 0.9532 & 0.9756 & 0.9872  \\ 
\cline{2-7}
&  VGGNet (UD+GS)  &  \textbf{0.7946} &  \textbf{0.9278} &  \textbf{0.9932} & \textbf{0.9969} & \textbf{0.9986}  \\
\cline{2-7}
&  VGGNet (UD+Softmax)  &  0.6389 &  0.8714 &  0.9798 & 0.9924 & 0.9974 \\
\cline{2-7}
&  VGGNet (SD+GS)  & 0.6564  & 0.9440 & 0.9812  & 0.9884 & 0.9906  \\
\cline{2-7}
&  VGGNet.CI (UD+GS) & 0.6242 & 0.8626 & 0.9745 & 0.9876 & 0.9946  \\
\hline

\multirow{5}*{head} 
& AlexNet (UD+GS) & 0.3888 & 0.7714 & 0.9419 & 0.9792 & 0.9886  \\
\cline{2-7}
&  VGGNet (UD+GS)  &  \textbf{0.7224} &  \textbf{0.9373} &  \textbf{0.9912} &  \textbf{0.9971} &  \textbf{0.9984}  \\
\cline{2-7}
&  VGGNet (UD+Softmax)  & 0.5893 &  0.8893 & 0.9828 & 0.9951 & 0.9976  \\
\cline{2-7}
&  VGGNet (SD+GS) & 0.3675 & 0.8267 & 0.9586  & 0.9848  & 0.9924\\
\cline{2-7}
& VGGNet.CI (UD+GS) & 0.5835 & 0.8766 & 0.9823 & 0.9938 & 0.9959  \\
\hline

\multirow{5}*{tooth} 
& AlexNet (UD+GS) & 0.6130 & 0.9038 & 0.9868 & 0.9958 & 0.9994 \\
\cline{2-7}
&  VGGNet (UD+GS)  & \textbf{0.8463} &  \textbf{0.9800} &  \textbf{0.9983} &  \textbf{0.9995} &  \textbf{0.9999}  \\
\cline{2-7}
&  VGGNet (UD+Softmax)  &  0.7291 &  0.9532 &  0.9954 & 0.9990 & 0.9997  \\
\cline{2-7}
&  VGGNet (SD+GS)  & 0.5963 & 0.9617  & 0.9898  & 0.9927 & 0.9932  \\
\cline{2-7}
& VGGNet.CI (UD+GS) & 0.6731 & 0.9259 & 0.9890 & 0.9984 & 0.9994  \\
\hline

\multirow{5}*{tree} 
& AlexNet (UD+GS) & 0.3417 & 0.7267 & 0.9268 & 0.9767 & 0.9918 \\
\cline{2-7}
&  VGGNet (UD+GS)  & \textbf{0.6679} &  \textbf{0.8896} &  \textbf{0.9882} &  \textbf{0.9970} &  \textbf{0.9992}  \\
\cline{2-7}
&  VGGNet (UD+Softmax)  & 0.5078 & 0.7922 & 0.9678 & 0.9930 & 0.9980  \\
\cline{2-7}
&  VGGNet (SD+GS)  & 0.3362  & 0.8108  & 0.9588  & 0.9863 & 0.9933  \\
\cline{2-7}
& VGGNet.CI (UD+GS) & 0.5015 & 0.8201 & 0.9585 & 0.9798 & 0.9921  \\
\hline

\multirow{5}*{vessel} 
& AlexNet (UD+GS) & 0.6355 & 0.9232 & 0.9950 & 0.9995 & 1.0000 \\
\cline{2-7}
&  VGGNet (UD+GS)  &  \textbf{0.9495} &  \textbf{0.9957} &  \textbf{0.9998} &  \textbf{1.0000} &  \textbf{1.0000} \\
\cline{2-7}
&  VGGNet (UD+Softmax) & 0.7948 & 0.9692 & 0.9975 & 0.9998 & 1.0000 \\
\cline{2-7}
&  VGGNet (SD+GS) & 0.5758 & 0.9637 & 0.9953  & 0.9970 & 0.9973 \\
\cline{2-7}
& VGGNet.CI (UD+GS) & 0.7130 & 0.9060 & 0.9725 & 0.9943 & 0.9984  \\
\hline
\end{tabular}
\end{table}
}

{\fontsize{9pt}{9pt}\selectfont
\begin{table}
\newcommand{\tabincell}[2]{\begin{tabular}{@{}#1@{}}#2\end{tabular}}
\caption{ Classification accuracy comparison of VGGNet (UD+GS), AlexNet (SD+GS) ~\cite{Su:2015:RCV} and $\mathcal{M}_G+$~\cite{Mahendran2018AMC} on the six categories under different tolerances.}
\begin{tabular}{|c|c|c|c|c|c|c|}
\hline
Cat. & Angle Tol. & $2^{\circ}$ & $5^{\circ}$ & $8^{\circ}$ & $11^{\circ}$ & $15^{\circ}$  \\
\hline
\multirow{3}*{engine}   
~ &  VGGNet (UD+GS) & \textbf{0.8450} & \textbf{0.9774} &  \textbf{0.9987} &  \textbf{0.9997} & \textbf{0.9999}  \\
\cline{2-7}
~ & AlexNet (SD+GS)~\cite{Su:2015:RCV} & 0.1571 & 0.5084 & 0.7669 & 0.8764 & 0.9185 \\
 \cline{2-7} 
~ & $\mathcal{M}_G+$~\cite{Mahendran2018AMC}  & 0.6016 & 0.9060 & 0.9672 & 0.9837 & 0.9920 \\
\hline

\multirow{3}*{fish} 
&  VGGNet (UD+GS)  &  \textbf{0.7946} &  \textbf{0.9278} &  \textbf{0.9932} & \textbf{0.9969} & \textbf{0.9986}  \\
\cline{2-7}
~ & AlexNet (SD+GS)~\cite{Su:2015:RCV} & 0.1361 & 0.4310 & 0.6454 & 0.7595 & 0.8215 \\
\cline{2-7}
~ &  $\mathcal{M}_G+$~\cite{Mahendran2018AMC}  & 0.3772 & 0.7883 & 0.9338 & 0.9737 & 0.9891 \\
\hline

\multirow{3}*{head} 
&  VGGNet (UD+GS)  &  \textbf{0.7224} &  \textbf{0.9373} &  \textbf{0.9912} &  \textbf{0.9971} &  \textbf{0.9984}  \\
\cline{2-7}
~ & AlexNet (SD+GS)~\cite{Su:2015:RCV} & 0.0691 & 0.2703 & 0.5121 & 0.6969 & 0.8136  \\
\cline{2-7}
~ & $\mathcal{M}_G+$~\cite{Mahendran2018AMC}  & 0.3018 & 0.6649 & 0.8655 & 0.9427 & 0.9699 \\
\hline

\multirow{3}*{tooth} 
&  VGGNet (UD+GS)  & \textbf{0.8463} &  \textbf{0.9800} &  \textbf{0.9983} &  \textbf{0.9995} &  \textbf{0.9999}  \\
\cline{2-7}
~ & AlexNet (SD+GS)~\cite{Su:2015:RCV} & 0.2123 & 0.6261 & 0.8770 & 0.9547 & 0.9759  \\
\cline{2-7}
~ &  $\mathcal{M}_G+$~\cite{Mahendran2018AMC}  & 0.4629 & 0.8191 & 0.9368 & 0.9721 & 0.9860 \\
\hline

\multirow{3}*{tree} 
&  VGGNet (UD+GS)  & \textbf{0.6679} &  \textbf{0.8896} &  \textbf{0.9882} &  \textbf{0.9970} &  \textbf{0.9992}  \\
\cline{2-7}
~ & AlexNet (SD+GS)~\cite{Su:2015:RCV} & 0.0748 & 0.2915 & 0.5344 & 0.7131 & 0.8195 \\
\cline{2-7}
~ &  $\mathcal{M}_G+$~\cite{Mahendran2018AMC}  & 0.2444 & 0.6264 & 0.8354 & 0.9319 & 0.9674 \\
\hline

\multirow{3}*{vessel} 
&  VGGNet (UD+GS)  &  \textbf{0.9495} &  \textbf{0.9957} &  \textbf{0.9998} &  \textbf{1.0000} &  \textbf{1.0000} \\
\cline{2-7}
~ & AlexNet (SD+GS)~\cite{Su:2015:RCV} & 0.2322 & 0.6698 & 0.8988 & 0.9707 & 0.9965 \\
\cline{2-7}
~ &  $\mathcal{M}_G+$~\cite{Mahendran2018AMC}  & 0.3600 & 0.7787 & 0.9335 & 0.9742 & 0.9888 \\
\hline
\end{tabular}
\end{table}
}

The classification accuracy of the testing dataset is shown in Table 2. For these rendered images, our model can obtain a good performance. First, under the 19-layer VGGNet, we compare our method (uniform division of the viewing sphere and geometric structure-aware loss function) with UD+Softmax (uniform division of the viewing sphere and Softmax loss function) and SD+GS (separate division of the azimuth and elevation and geometric structure-aware loss function) on the six categories under different tolerances. 
The same training dataset is used to train these classification models, and the classification accuracy under different tolerances is also listed in Table 2. The models using uniform division of the viewing sphere (UD+GS and UD+Softmax) have better performance than the model with a separate division of the azimuth and elevation (SD+GS). The reason is that when dividing the viewing sphere into uniform regions, the CNN can optimize a more straightforward problem by the geodesic distance, instead of the angle difference in the azimuth and elevation. Besides, the proposed geometric structure-aware loss function is better than the Softmax loss function. The comparison between VGGNet (UD+GS) and AlexNet (UD+GS) shows the effect of a deeper network, i.e., the 19-layer VGGNet. VGGNet has better performance than AlexNet. Furthermore, we compare our category-specific model with the category-independent network. In the category independent network, all the convolution layers and fully connected layers except the last one are shared by all classes, while class-dependent layers (one fc layer for each class) are stacked over them. The category-independent network can save parameters for the whole system and have similar performance to VGGNet (UD+Softmax), but it would reduce the system's prediction accuracy for each category, as shown in Table 2. Taking all the comparison into consideration, we choose the VGGNet (UD+GS) model for the following sections (Error Analysis and Applications). 

We further compare our VGGNet (UD+GS) model with two state-of-the-art methods, AlexNet (SD+GS)~\cite{Su:2015:RCV} and $\mathcal{M}_G+$ (Geodesic Bin \& Delta Mode) \cite{Mahendran2018AMC} on the six categories under different tolerances. For AlexNet (SD+GS), in our experiment, we ignore the camera-lit angle and only care about the azimuth and elevation, as we did for VGGNet (SD+GS). $\mathcal{M}_G+$ model predicts the viewpoint label first by classification, then estimates the viewpoint residual by regression, and finally combines the viewpoint label and residual to obtain the final viewpoint. In the experiment, we choose the size of the K-means dictionary $K = 24$ and the importance of the geodesic distance $\alpha = 10$,  under which the model achieves the best performance. A modification has been made to these methods, namely, the viewpoint estimation network is now category-specific rather than category-independent in their original experiments, since category-specific networks are proved better in previous experiments. As shown in Table 3, our result is better than the result of AlexNet (SD+GS), VGGNet (SD+GS) and $\mathcal{M}_G+$ model. 
Su et al.~\cite{Su:2015:RCV} and Mahendran et al.~\cite{Mahendran2018AMC} employed a large angle tolerance ($30^{\circ}$) for viewpoint estimation of 3D models. Our angle tolerance is much less than $30^{\circ}$, and this indicates that our method is relatively more accurate for viewpoint estimation for volumes and the estimated viewpoints can be used in the following applications.

\subsection{Error Analysis}

As shown in Table 2, for the VGGNet (UD+GS) model, except for the vessel category, the angle differences of some of the rendered images are larger than $5^{\circ}$. Thus, we analyzed which kinds of rendered images or features are hard to estimate their viewpoints. We count the number of misclassified images under Acc-$5^{\circ}$ as the classification error for each ground-truth viewpoint. This results in an error map on a 2D azimuth-elevation plane.

\begin{figure}
  \centering
  \includegraphics[width=0.75\linewidth]{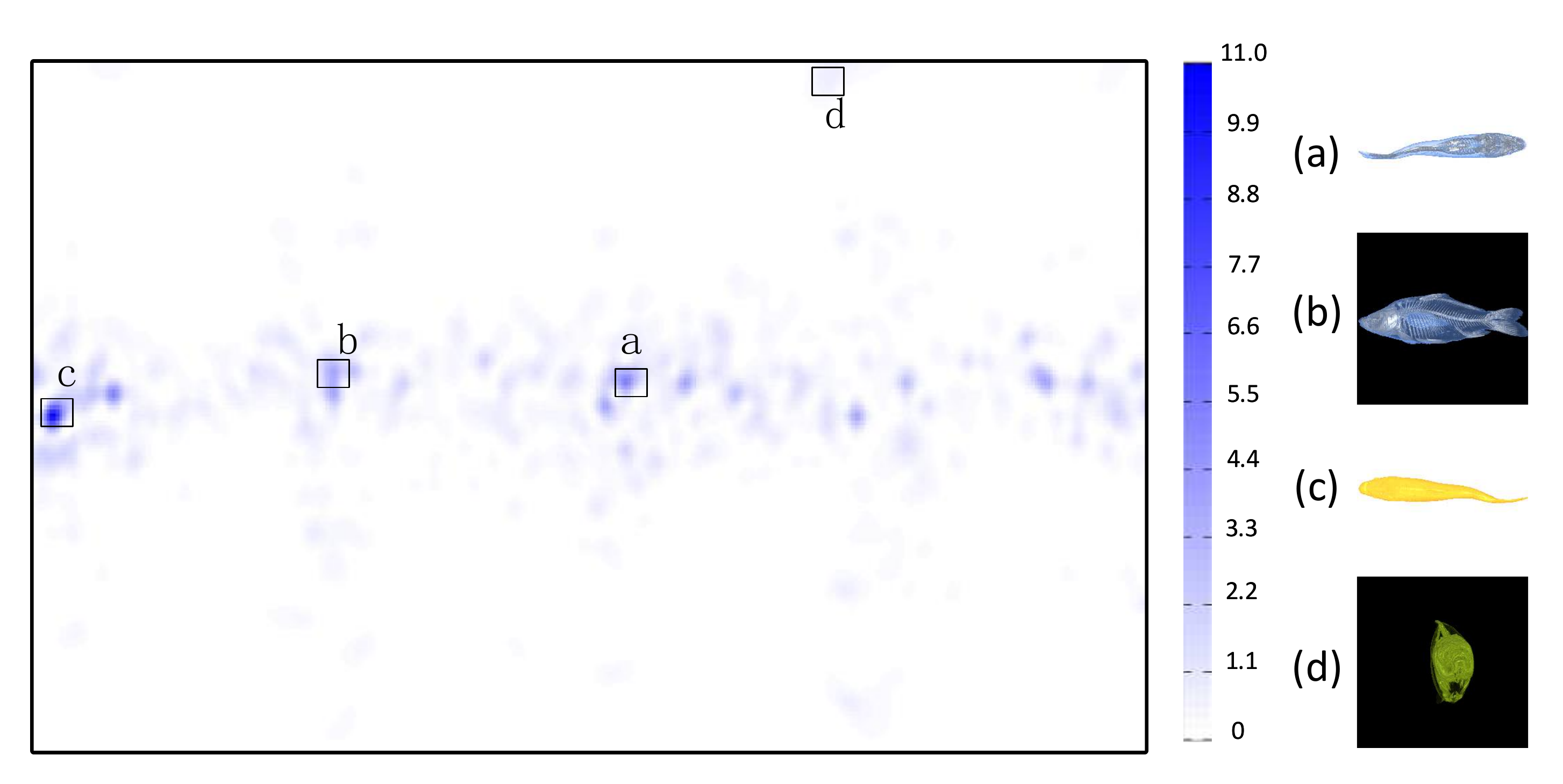}
  \caption{The classification error map for the fish category. The viewpoints in the blue region are more likely to be misclassified. (a) a side view, (b) a front view, (c) another side view, and (d) a top view from the head.}
  \label{fig:fishErrorAnalysis}epstopdf --outfile=pipeline-eps-converted-to.pdf pipeline.eps
\end{figure}

Fig.~\ref{fig:fishErrorAnalysis} is the error map for the fish category. At the front and the side views (Fig.~\ref{fig:fishErrorAnalysis}(a)-(c)), the classification error is a little higher than most other well-classified viewpoints, although the rendered image of the estimated viewpoint is very similar to the one of the ground-truth viewpoint. The relative higher classification error can be explained by the view stability~\cite{Bordoloi:2005:VSV} of these viewing regions, which means a small change occurs when the camera is shifted within a small neighborhood.

Furthermore, the rendered images under the top view and bottom view (Fig.~\ref{fig:fishErrorAnalysis}(d)) are likely to be misclassified into symmetrical viewpoints. This is due to that their outer contours being similar to each other, but the inner features are not clear enough due to visual clutter. There are no significant gradient changes in the rendered image, so that it is confusing for our viewpoint estimation network to identify the right viewpoint.

\begin{figure}
  \centering
  \includegraphics[width=0.75\linewidth]{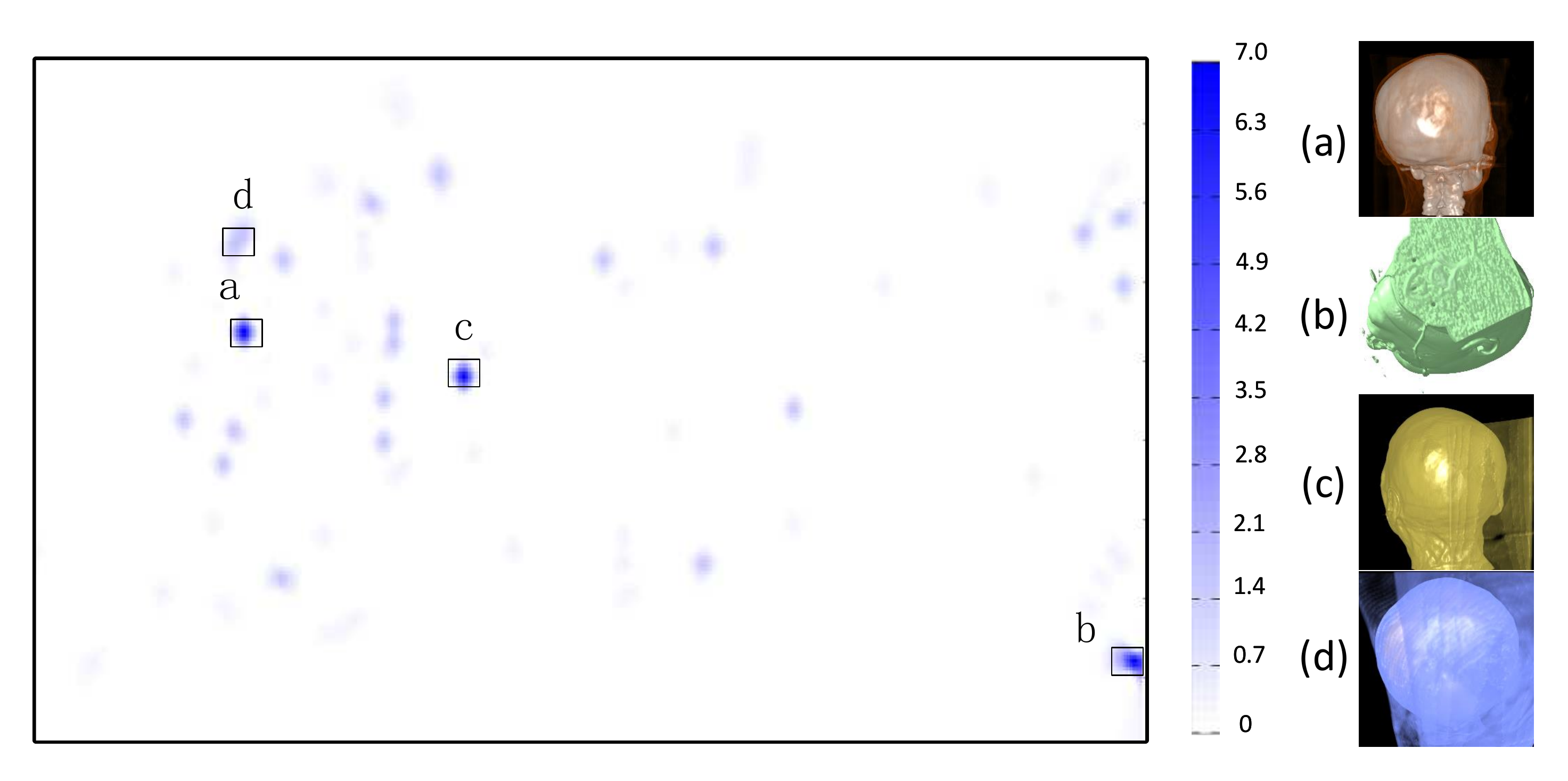}
  \caption{The classification error map for the head category. The viewpoints in the blue region are more likely to be misclassified. (a)-(d) are four rendered images at representative misclassified viewpoints. }
  \label{fig:headErrorAnalysis}
\end{figure}

We also analyzed the classification error of the head category and four misclassified examples, as shown in Fig.~\ref{fig:headErrorAnalysis}. The misclassified viewpoints are generally distributed among the back, up and bottom views. These images generally have a lack of distinguishable features, and the viewpoints in these regions are relatively stable. Thus, it is hard to distinguish them from nearby viewpoints due to their featureless rendered images. In the front view, since our network can identify rich facial features, the classification accuracy is relatively high.

\begin{figure}
  \centering
  \includegraphics[width=1\linewidth]{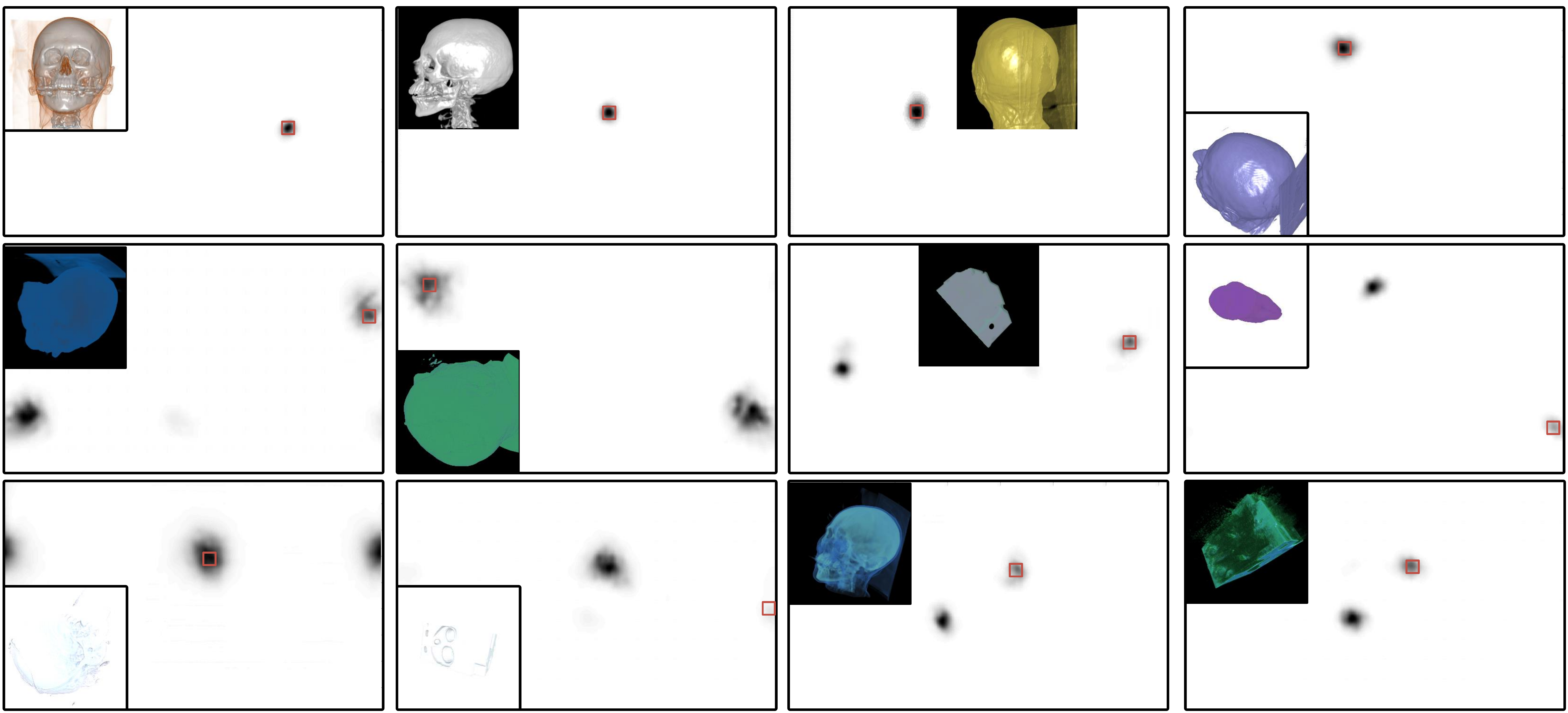}
  \caption{The classification probability distributions of 12 representative images. On the viewing sphere in 2D, our method classifies the rendered image to the viewpoints in the black region with a high probability. The red  box indicates the ground-truth viewpoint. The first row has only one bivariate Gaussion distribution. The middle and bottom rows have two bivariate Gaussian distributions, and the lower peak is around the ground-truth viewpoint.}
  \label{fig:badClassification}
\end{figure}


It is worth noting that some rendered images are obviously misclassified, especially, those misclassified under the 50$^{\circ}$ tolerance. Fig.~\ref{fig:badClassification} shows eight representative examples. We are interested in what their classification probability distributions look like and why these images are misclassified. We observe an interesting pattern from our representative examples in Fig.~\ref{fig:badClassification}. 
Although the geodesic distance between the highest peak produced by our method and the ground-truth viewpoint is not within the tolerance, there is still a lower peak around the ground-truth viewpoint. The lower peak can contribute to viewpoint selection in our applications, and this can only be achieved with the classification model, instead of the regression model. 

These misclassifications in Fig.~\ref{fig:badClassification} are due to unrecognizable internal features. We observe some typical misclassification patterns in our results: bad light conditions and bad opacity transfer functions. In the case of the environment light only, some rendered images do not have enough recognizable features for our model. This phenomenon also occurs when the light intensity is too high. In case of a bad opacity transfer function, it may lead to visual clutter in the rendered image. This usually happens when the opacity is low. Thus, when the inner structure of the volume is complex, such as the Chapel Hill CT Head and the engine, its inner structure will be mixed together due to its transparency. As a result, the rendered image is very similar to the one under a symmetrical viewpoint on the viewing sphere. Thus, these images are likely to be misclassified into symmetrical viewpoints, and this results in an ambiguity of viewpoint estimation. When the opacity is high for the outer feature as the context, it may occlude important inner features, which also makes our model confused and results in the ambiguity of viewpoint estimation.

In summary, some viewpoints are hard to estimate, and this misclassification may be due to the high image similarity between the estimated viewpoint and the ground-truth viewpoint. The rendered images of stable viewpoints are similar to those of nearby viewpoints, and transparent structures may have the same rendered image from the viewpoint and the symmetric viewpoint. In addition, the featureless images are also less distinguishable.

\section{Applications}

Our viewpoint estimation method can be used to support a variety of volume visualization applications,
such as extending the transfer function exploration from a rendered image~\cite{Raji:2017:PGE}. In this
section, we describe two direct applications of viewpoint estimation: viewpoint preference analysis and viewpoint selection based on CNN-feature similarity. we describe two direct applications of viewpoint estimation based on the VGGNet (UD+GS) model.

\subsection{Viewpoint Preference Analysis}


There are many rendered images in published papers on volume visualization. Most viewpoints for these
images are selected by visualization or domain experts to maximize the amount of information about features
in the rendered image or to highlight important features. Their viewpoint preferences are expressed in
these rendered images. Thus, we can apply our viewpoint estimation method to analyze their preferences.

We utilize the image database~\cite{Tao:2016:SVV}, collected from visualization journals and conferences.
The categories are the same with the one in Table 1, and our trained category-specific CNNs can be used to
recover the viewpoints from collected images. Since the ground-truth viewpoints of collected images are unknown, 
we visually compared the rendered images of recovered viewpoints
with collected images, and most of them can be classified correctly, with little difference between rendered
and collected images. For collected images whose source volume is beyond our training dataset, most of them can also be estimated correctly. Some collected images are misclassified, due to various reasons, such as different
projection type and non-uniform background color. Considering the distribution of the estimated viewpoint
$N(v_\mu,v_\sigma^2)$, we find that $v_\sigma$ is very high for misclassified images, which means the
probability of the viewpoint $v_\mu$ is relatively small.


Since most collected images generate a bivariate Gaussian probability distribution on viewpoints, we
accumulate the distribution of each collected image in the category to generate a viewing map to analyze
the viewpoint preference for the volume in this category. The distribution in the viewing map is similar to the
Gaussian mixture model. Due to the small probability of misclassified images, their influence on viewpoint
preference analysis is relatively limited.

\begin{figure}
  \centering
  \includegraphics[width=0.9\linewidth]{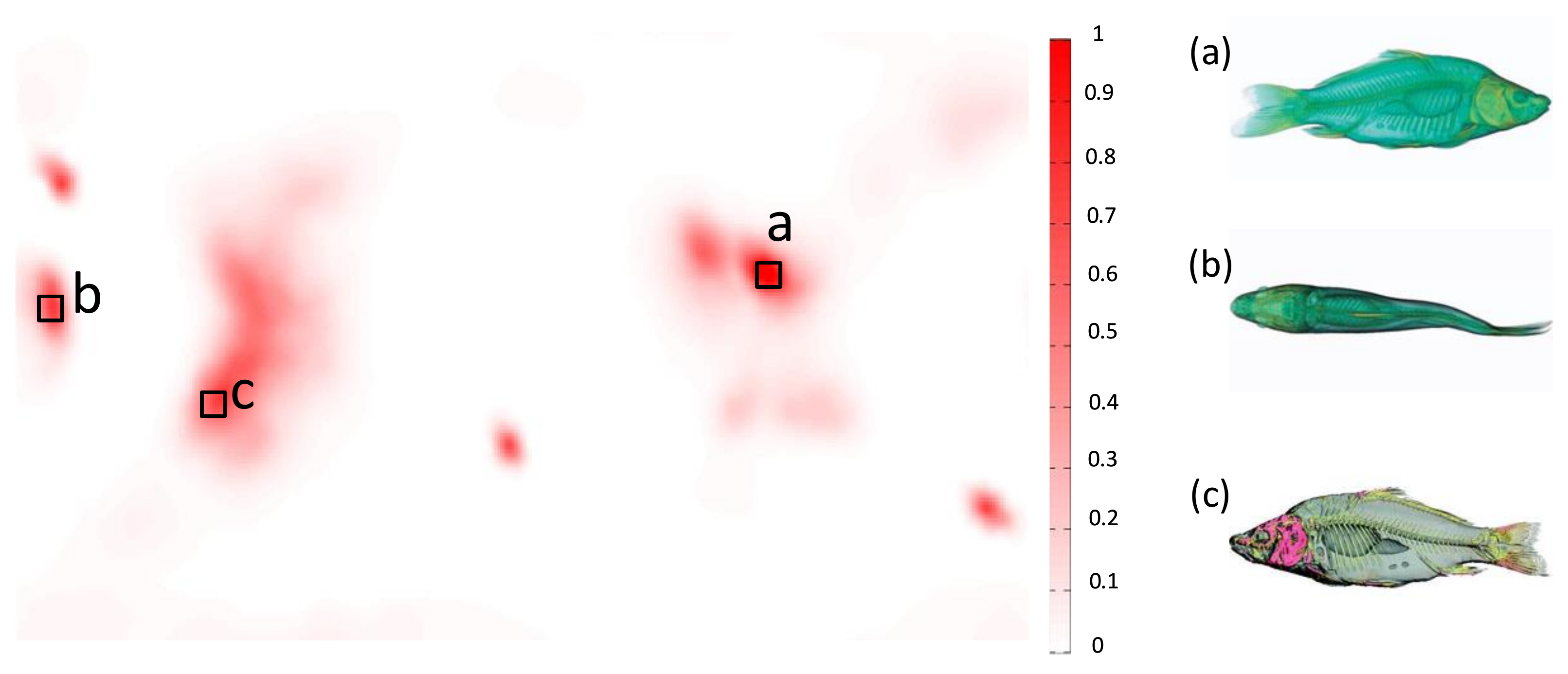}
  \caption{The viewing map shows the estimated viewpoints of collected images in the fish category. Experts usually have viewpoint preferences for the viewpoints in the red region. (a)-(c) are three collected images at representative viewpoints labeled in the viewing map.}
  \label{fig:fishmap}
\end{figure}

Fig.~\ref{fig:fishmap} shows the viewing map of the fish category. There are three regions with a high
probability. Experts tend to select side viewpoints to avoid occlusions. In addition, there is one viewpoint cluster at the fish's back, revealing the swimming pose of the fish.

\begin{figure*}
  \centering
  \includegraphics[width=1\linewidth]{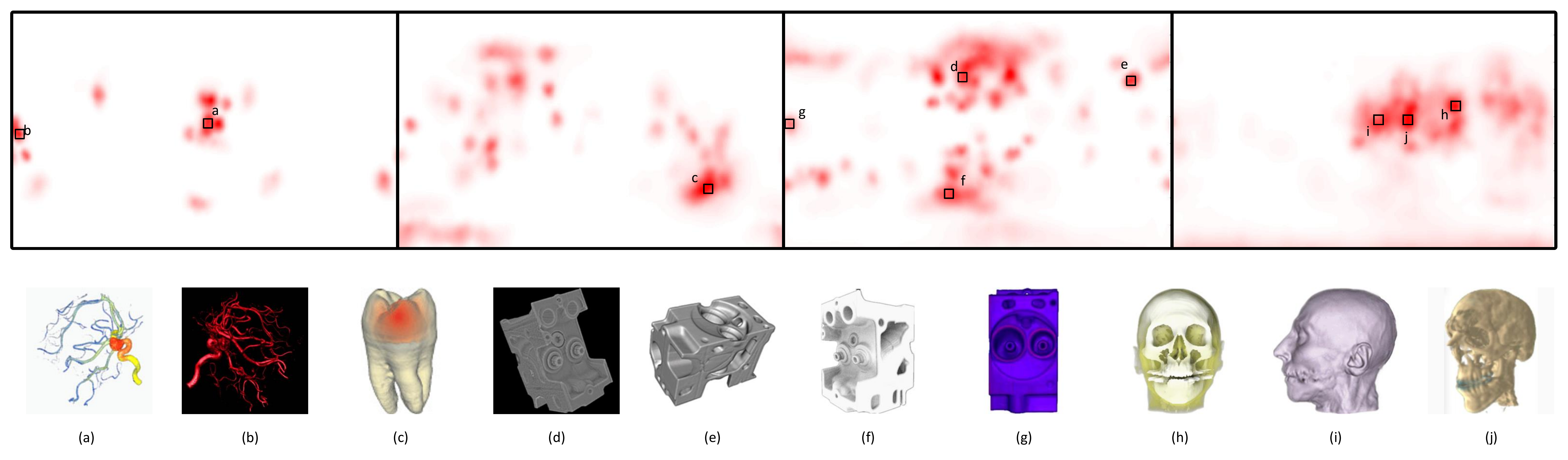}
  \caption{The viewing map shows the estimated viewpoints of collected images in the foot, tooth, engine, and head categories. Experts usually have viewpoint preferences for the viewpoints in the red regions. (a)-(j) are collected images at representative viewpoints.}
  \label{fig:viewingmap}
\end{figure*}

We also analyze the other four categories: the vessel, tooth, engine, and head categories in Fig.~\ref{fig:viewingmap}.
According to the probability distribution estimated from the collected images of the vessel category, experts tend to
display the volume in the front view (Fig.~\ref{fig:viewingmap}(a)), since they are more interested in the aneurism and would like to avoid any occlusions. Many experts' viewpoint preference for the tooth category would generally focused on the lower right corner of the viewing map, while there are also other viewpoints for the tooth. It is necessary to explain that some images are misclassified to symmetrical viewpoints because they are mirrored, and some images in the image dataset are ``worst case'' viewpoints. In contrast, the viewpoints are quite diverse for the engine category. As the structure of the engine is quite complex and different viewpoints can reveal different structures, experts use different viewpoints to comprehensively understand the engine. They tend to select the three-quarter views (Fig.~\ref{fig:viewingmap}(d)) and also choose side views (Fig.~\ref{fig:viewingmap}(g)) as supplements. The preferred viewpoints in the head category are clustered in the front view, side view and three-quarter view (Fig.~\ref{fig:viewingmap}(h-j)). There is no clear boundary between these regions. This is due to different experts have slightly different preferences from the front view to the side view.

For the viewpoint preference analysis application, we need only the trained CNN and collected images in this
category. The analysis result can tell us how users select the viewpoint for the volume in this category,
and find features interesting to most users.

\subsection{Viewpoint Selection based on CNN-feature similarity}

Given an input volume and a transfer function, the viewpoint selection application suggest the optimal
viewpoint for features of interest. The viewpoint preference analysis does not consider the input volume, and suggests the same representative viewpoint for different features of the same volume and different volumes. It would be better to consider the similarity between the input features and the features in the collected images. Inspired by the similarity voting for viewpoint selection~\cite{Tao:2016:SVV}, we propose a weighted probability voting for viewpoint selection.

\begin{figure}
  \centering
  \includegraphics[width=1\linewidth]{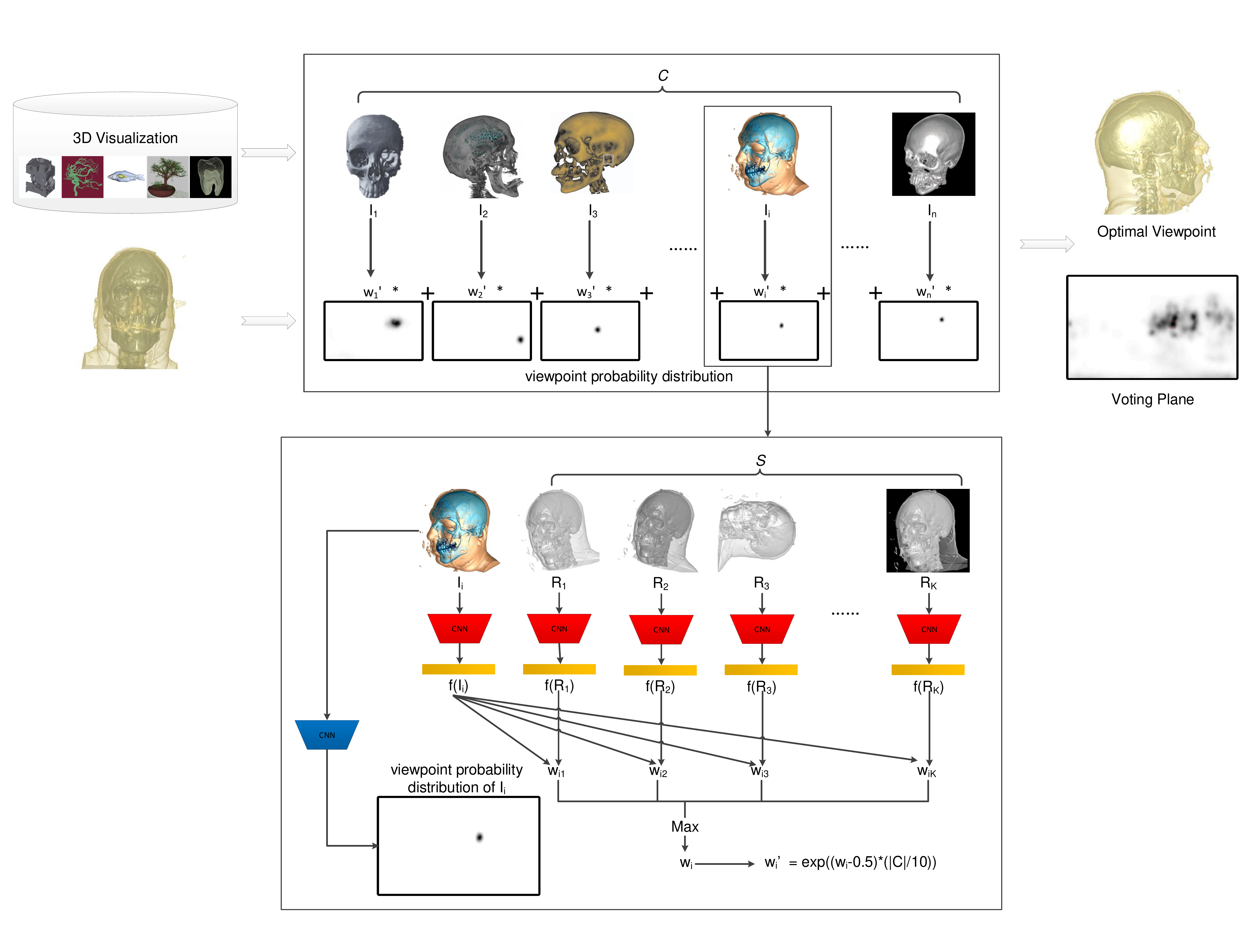}
  \caption{CNN-feature based viewpoint selection pipeline. The first stage is CNN-feature based similarity computation. The input volume is rendered with the provided transfer function and the estimated viewpoint of the collected image $I_i$ to generate a rendered image set $S$. The similarity $w_i$ between the input volume and  $I_i$ can be computed with the help of the features extracted by  the category classification network. Then the viewpoint probability distribution of $I_i$ is weighted by the similarity $ {w_i}'$ to generate the voting probability distribution of $I_i$. Finally, all weighted voting probability distributions are summed up, and the viewpoint with the largest probability corresponds to the optimal viewpoint.}
  \label{fig:similarity}
\end{figure}

We first need to determine the input volume belonging to which category. AlexNet 
is used to train our category classification network. The volume is classified into seven categories, six of which are listed in Table 1, and the last category is \emph{others}, not belonging to the six categories. This network is trained through the same training dataset with additional images of other volumes. For the input volume, we first render several images randomly using the provided transfer function, and classify these images
through the category classification network. The most voted category is considered the category of the volume.

After obtaining the category of the input volume, the weighted probability voting is illustrated in Fig.~\ref{fig:similarity}. We denote the collected image set of this category as $C = \{I_1, I_2, \ldots, I_n\}$, where $n$ is the number of collected images. For each collected image $I_i \in C$, we can estimate its viewpoint $v_i$ with our viewpoint estimation network. We then render the volume with the provided transfer function under the estimated viewpoint $v_i$ considering different camera-tilt angles and background colors. Thus, we have $ k=40$ images denoted as the rendered image set $S = \{R_1, R_2, \ldots, R_k\}$.
We then calculate the similarity between the rendered image set and the collected image $I_i$ as the voting weight for the collected image $I_i$.
Since the category classification network extracts features for volume classification, we can employ the features of the last hidden layer when classifying the rendered image $R_j\in S$ and the collected image $I_i$, denoted by $f(R_j)$ and $f(I_i) \in R^{4096}$, to measure the similarity $w_i$ between the collected image $I_i$ and the input volume together with the transfer function using the cosine distance as follows
\begin{equation}
\max_{R_j \in S}(cos(f(R_j), f(I_i))).
\end{equation}

For each collected image, its probability is weighted by its similarity with the input volume and transfer function,
and we can further design the exponential similarity by $ {w_i}' = exp((w_i - 0.5)*|C|/10) $, where $|C|$ is number of images in the collected image set. All weighted
probability distributions are summed up, and the viewpoint with the largest probability corresponds to the optimal viewpoint.

\begin{figure}
  \centering
  \subfigure[] {\label{fig:engine}   \includegraphics[width=0.2\linewidth]{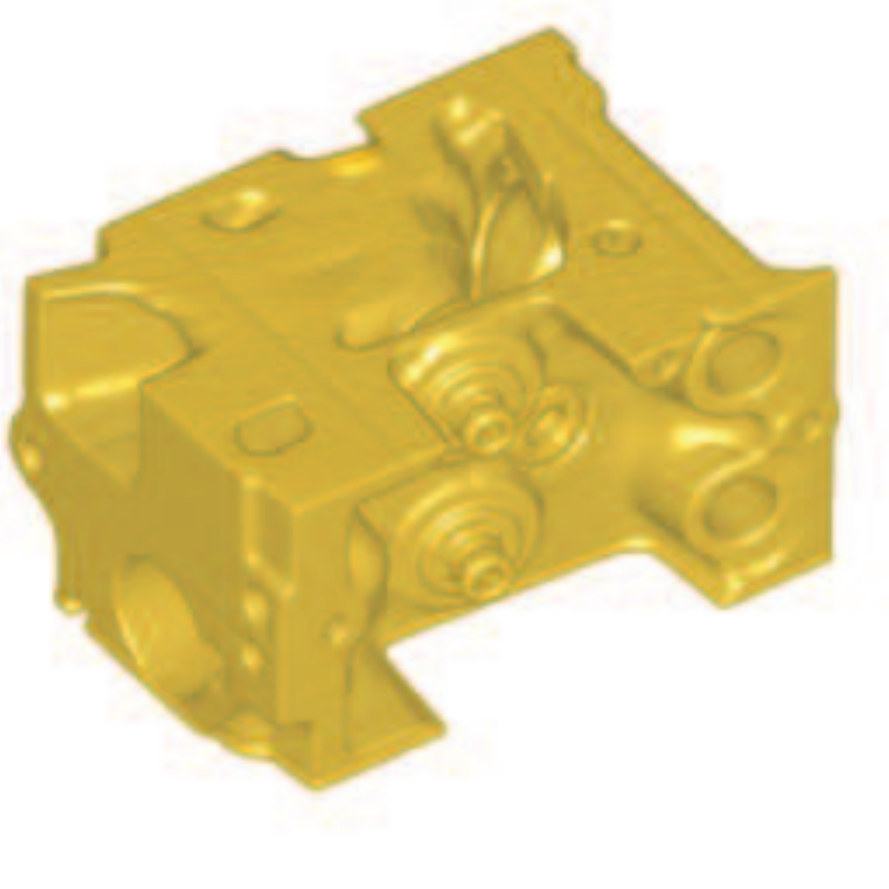}}
  \subfigure[] {\label{fig:vessel}   \includegraphics[width=0.2\linewidth]{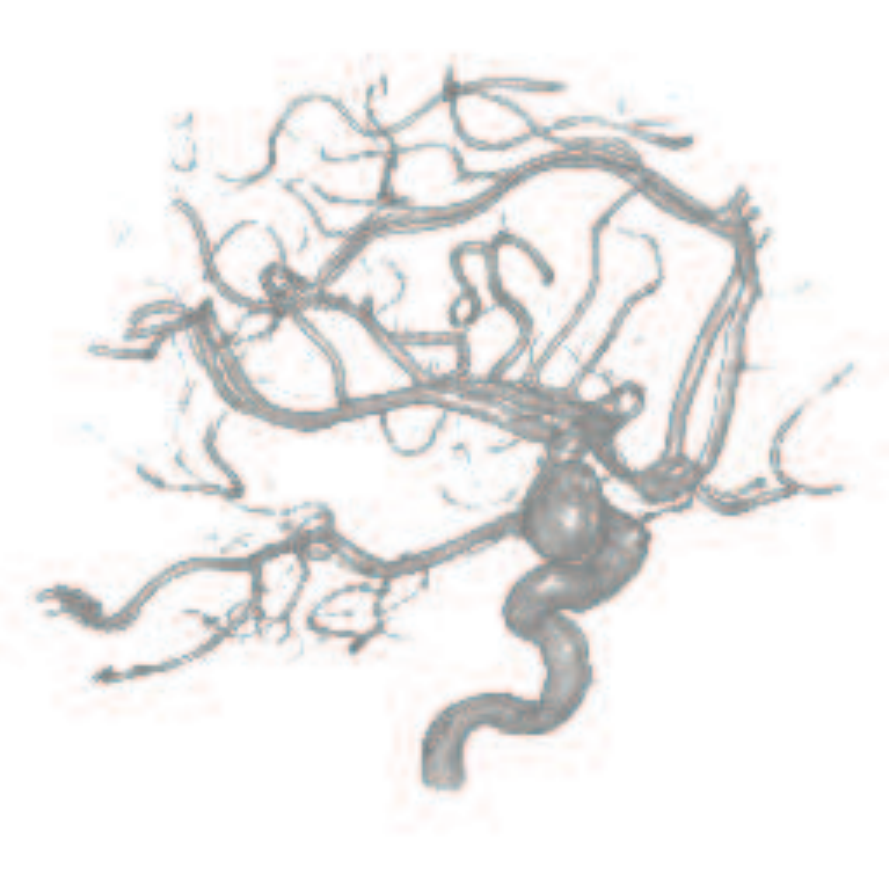}}
  \subfigure[] {\label{fig:tree2113} \includegraphics[width=0.2\linewidth]{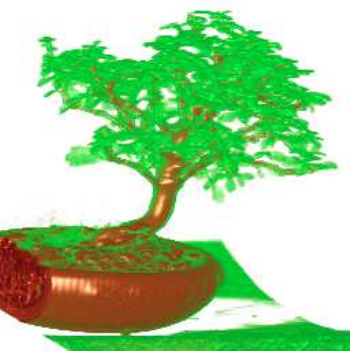}}
  \caption{The viewpoint selection results for the engine, the vessel and the bonsai tree from (a) to (c), respectively. }
  \label{fig:optimalviewpoints}
\end{figure}

We first evaluate our method by three volumes: the engine, the vessel and the bonsai tree, as shown in Fig.~\ref{fig:optimalviewpoints}. The optimal viewpoint for the engine is quite close to the three-quater view, which is preferred by a lot of users. For the vessel, since users are generally more interested in the aneurism, our selected viewpoint avoids the occlusion on it and shows the aneurism clearly. For the bonsai tree, our result is not only concerned with the clearness of the  semantically important features, such as the trunk, tree branches and leaves, but it also shows other meaningful features, such as the soil, the grass and the base plane by a lightly oblique shift.  

\begin{figure*}[tbp]
  \centering
  \subfigure[] {\label{fig:ctbone}  	   \includegraphics[width=0.21\linewidth]{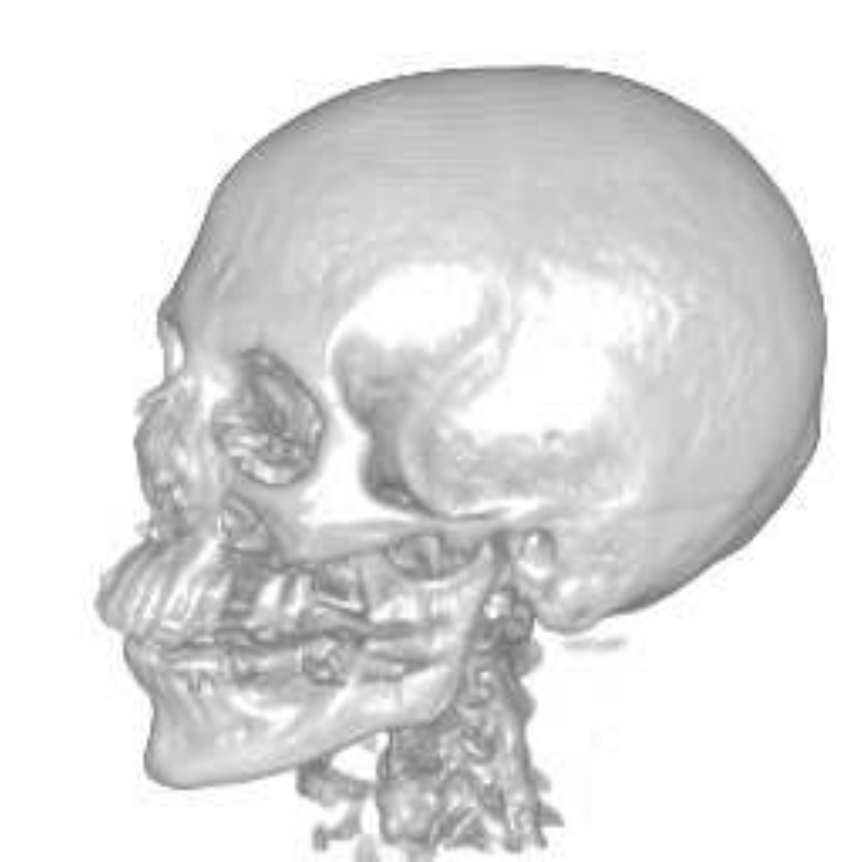}}
  \subfigure[] {\label{fig:mriskine}        \includegraphics[width=0.21\linewidth]{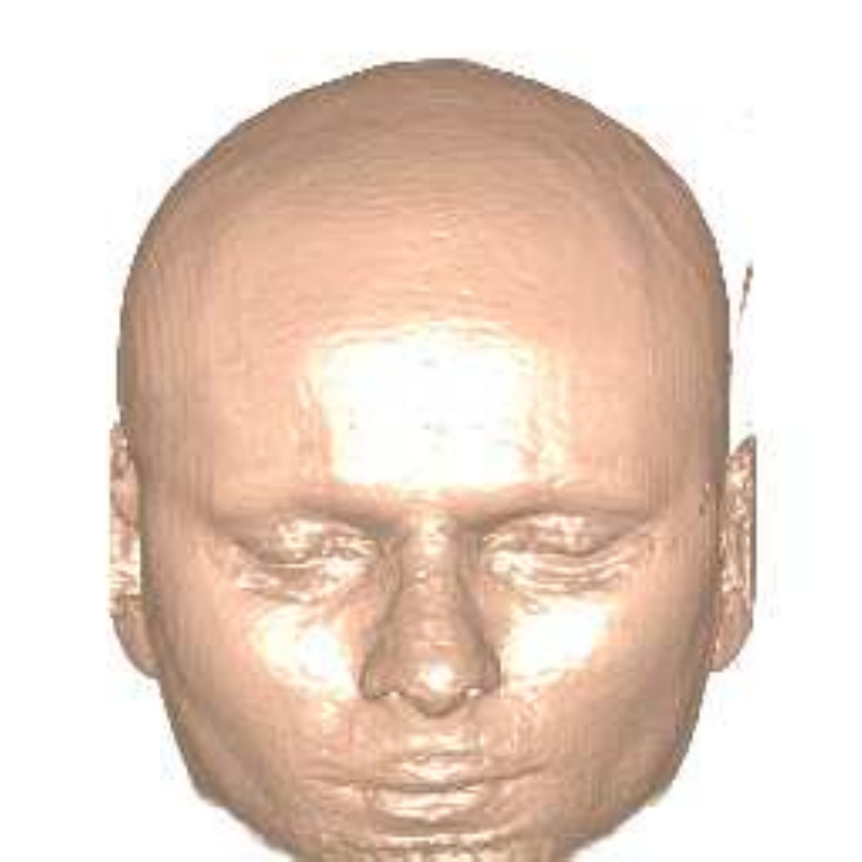}}
  \subfigure[] {\label{fig:vismalemixture} \includegraphics[width=0.21\linewidth]{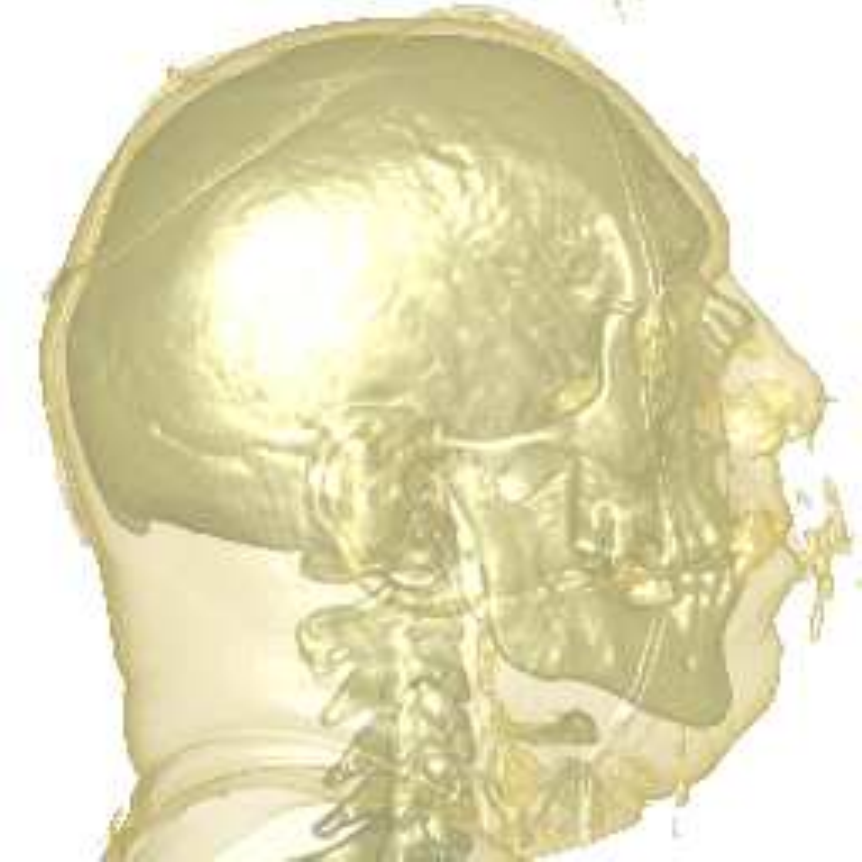}}
  \subfigure[] {\label{fig:vismaleskin}    \includegraphics[width=0.21\linewidth]{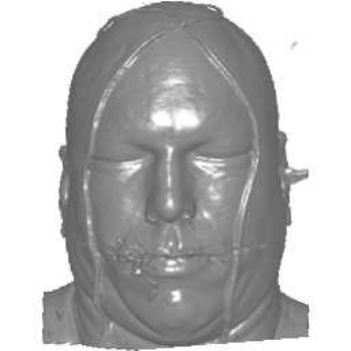}}
  \caption{The viewpoint selection results of the heads. (a) The optimal viewpoint for the bone of the Chapel Hill CT head. 
	      (b) The optimal viewpoint for the skin of the MRI head. (c) The optimal viewpoint for the bone and skin of the visual male. (d) The optimal viewpoint
          for the skin of the visual male.}
  \label{fig:head}
\end{figure*}

Our method can suggest different optimal viewpoints for different features of the same volume and different volumes. Since the head category has three different volumes and each volume has different features. 
Our method is used to choose the optimal viewpoint, in particular, for currently visible features in the head category, as shown
in Fig.~\ref{fig:head}. For the MRI head, we can clearly observe the facial features from the front
view, thus the front view is regarded as the optimal viewpoint in Fig.~\ref{fig:mriskine}. It is the same for the head of the visual male with only the skin in Fig.~\ref{fig:vismaleskin}. However, for the Chapel Hill CT head with the bone, the side view is selected to show more structures of the bone in Fig.~\ref{fig:ctbone}. For the head of the visual male with the bone and the skin, the front view shows visual clutter, and we can display clearer features on the side view. In this situation, our method suggests the side view as the optimal viewpoint in Fig.~\ref{fig:vismalemixture}.

The image-based viewpoint selection model~\cite{Tao:2016:SVV} applies one single similarity measure to viewpoint selection, and mixes two separate stages: viewpoint estimation and similarity calculation. Our method separates these two stages and solves them by CNNs. In the following, we compare our method with the image-based viewpoint viewpoint selection model by the viewpoint estimation accuracy and viewpoint selection result.



As the viewpoints of the collected images in published papers are unknown, we randomly generate some viewpoint-annotated images as the collected images for each category for viewpoint estimation evaluation. 
In~\cite{Tao:2016:SVV}, every collected image votes on the 
viewpoints of its 12  most similar images with the same weight when the number of viewpoints is 2,352. Thus, the similarity between the voted viewpoints and the ground-truth viewpoint is 
the key to viewpoint selection. We apply Acc-$n^{\circ}$ as the average probability of the 12 selected viewpoints within the $n^{\circ}$ neighbor region of the ground-truth viewpoint,  instead of the probability of the optimal estimated viewpoint in Section 4.3.
For each collected image, we evaluate the 12 voted viewpoints under Acc-19$^{\circ}$, Acc-22$^{\circ}$, Acc-25$^{\circ}$ and Acc-29$^{\circ}$, and the average accuracy for images in each category is shown in Table 4. 
Since SIFT and HOG are designed primarily for image classification and object detection, the performance on viewpoint estimation of randomly generated images is 
not very good, especially for the image without too many features. For further validation, we manually generate several rendered images under the viewpoints of  
images from published papers~\cite{Tao:2016:SVV}. The evaluation result in Table 5 is better than the one for randomly generated images in Table 4. However, compared with our estimation 
result for the 12 selected viewpoints Acc-8$^{\circ}$ in Table 4 and Table 5, our method has better performance on viewpoint estimation.

{\fontsize{9pt}{9pt}\selectfont
\begin{table}
\newcommand{\tabincell}[2]{\begin{tabular}{@{}#1@{}}#2\end{tabular}}
\caption{The viewpoint estimation accuracy of images with general viewpoints using the image-based viewpoint selection model~\cite{Tao:2016:SVV}. }
\begin{tabular}{|c|c|c|c|c|c|c|}
\hline
Category & Size & Acc-19$^{\circ}$ & Acc-22$^{\circ}$ & Acc-25$^{\circ}$ & Acc-29$^{\circ}$ &  Acc-8$^{\circ}$(Our)\\
\hline
engine & 450 & 0.354 & 0.383 & 0.406 & 0.422 & \textbf{0.870}\\
\hline
fish & 300 & 0.187 & 0.200 & 0.235 & 0.252 & \textbf{0.769} \\
\hline
head & 350 & 0.310 & 0.356 & 0.393 & 0.417 & \textbf{0.842} \\
\hline
tooth & 250 & 0.303 & 0.329 & 0.342 & 0.355 &  \textbf{0.871}\\
\hline
tree & 150 & 0.236 & 0.284 & 0.308 & 0.340 & \textbf{0.755}\\
\hline
vessel & 100 & 0.342 & 0.387 & 0.423 & 0.456 & \textbf{0.875}\\
\hline
\end{tabular}
\end{table}
}

{\fontsize{9pt}{9pt}\selectfont
\begin{table}
\newcommand{\tabincell}[2]{\begin{tabular}{@{}#1@{}}#2\end{tabular}}
\caption{The viewpoint estimation accuracy of images with manually selected representative viewpoints using the image-based viewpoint selection model~\cite{Tao:2016:SVV}. }
\begin{tabular}{|c|c|c|c|c|c|c|}
\hline
Category & Size & Acc-19$^{\circ}$ & Acc-22$^{\circ}$ & Acc-25$^{\circ}$ & Acc-29$^{\circ}$ &  Acc-8$^{\circ}$(Our) \\
\hline
engine & 513 & 0.487 & 0.527 & 0.555 & 0.575 & \textbf{0.859}\\
\hline
fish & 324 & 0.376 & 0.425 & 0.465 & 0.498 & \textbf{0.791}\\
\hline
head & 399 & 0.387 & 0.430 & 0.465 & 0.488 & \textbf{0.867}\\
\hline
tooth & 250 & 0.379 & 0.398 & 0.417 & 0.436 & \textbf{0.880}\\
\hline
tree & 150 & 0.330 & 0.368 & 0.400 & 0.430 & \textbf{0.727}\\
\hline
vessel & 100 & 0.464 & 0.518 & 0.548 & 0.569 & \textbf{0.898} \\
\hline
\end{tabular}
\end{table}
}

\begin{figure*}[tbp]
  \centering
  \subfigure[] {\label{fig:tooth-new}   \includegraphics[width=0.15\linewidth]{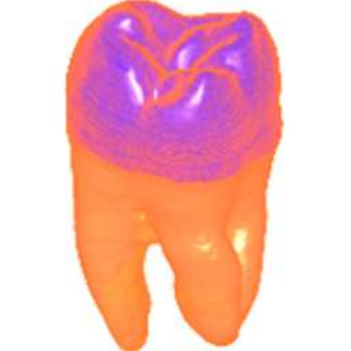}}
  \subfigure[] {\label{fig:tooth-old}   \includegraphics[width=0.15\linewidth]{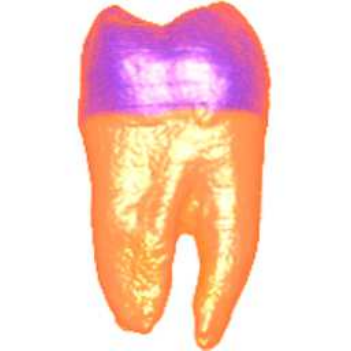}}
  \subfigure[] {\label{fig:mri-new} \includegraphics[width=0.15\linewidth]{mri1-1525}}
  \subfigure[] {\label{fig:mri-old}  \includegraphics[width=0.15\linewidth]{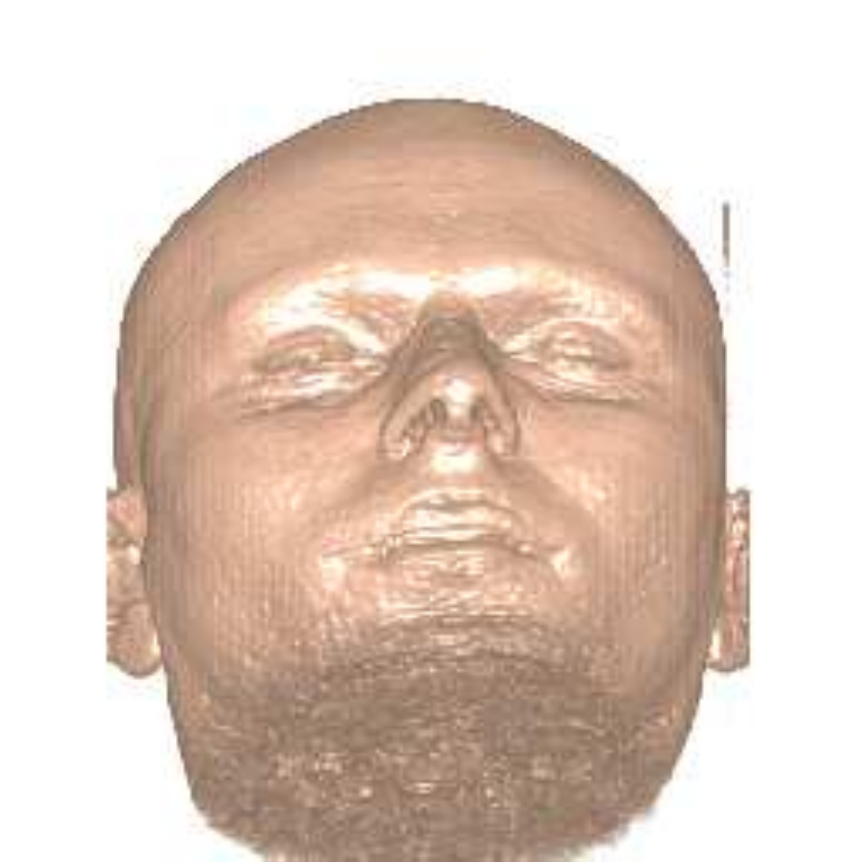}}
  \subfigure[] {\label{fig:tree-new}  \includegraphics[width=0.15\linewidth]{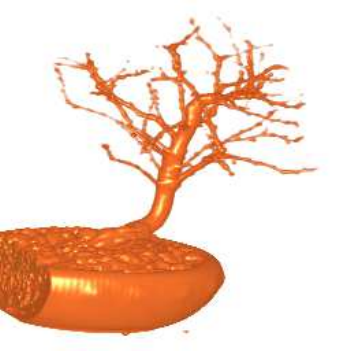}}
  \subfigure[] {\label{fig:new-old}  \includegraphics[width=0.15\linewidth]{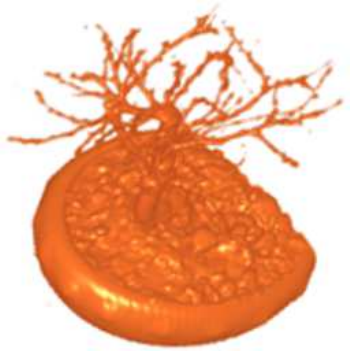}}
  \caption{Comparison of the optimal viewpoints for the tooth, the MRI head, and the bonsai tree by two methods: our method (left) and the similarity voting based viewpoint selection method~\cite{Tao:2016:SVV} (right). }
  \label{fig:compareviewpoints}
\end{figure*}

The viewpoint selection results of the tooth, the MRI head and the bonsai tree of our method and the image-based viewpoint selection model are shown in Fig.~\ref{fig:compareviewpoints}. All three volumes have clear semantic meanings, espcially the up direction. For the tooth volume, instead of a completely front view, our method selects the viewpoint with a slightly oblique shift to reveal the crown structure more clearly without occlusion. For the MRI head, both methods choose the front view, but the viewpoint of our method is more consistent with the aesthetic criterion. For the bonsai tree, our method can capture the up direction from collected images. However, the trunk, branch, and soil can not be easily separated under the viewpoint from the image-based viewpoint selection model. As a result, our CNN-feature based image similarity can suggest more semantically meaningful viewpoints than the SIFT and HOG based image similarity.

\section{Conclusion}
In this paper, we propose a CNN based viewpoint estimation method. Inspired by the ``Render for CNN'' approach, an overfit-resistant image rendering pipeline was designed to generate training images with viewpoint annotations considering different transfer functions and rendering parameters. These images are used to train a category-specific viewpoint classification network. The proposed method was tested on six categories based on available online volumes. We can achieve a classification accuracy of at least 0.89 in the maximum angle difference $5^\circ$. Our viewpoint estimation model is better than previous methods due to its better viewing sphere division and the geometric structure-aware loss function. We successfully applied our method on two applications: the viewpoint preference analysis of collected images in publications, and a CNN-feature similarity based viewpoint selection.

In the future, we would like to replace the manually-designed opacity transfer functions with image-driven or data-driven opacity transfer functions to improve the richness of features and the training efficiency when training a new category. When more volumes are available, we will add them as training volumes in the corresponding category to improve the generalization, especially for categories with only one volume. We also plan to recover the transfer function of a rendered image based on the estimated viewpoint, or jointly learn the viewpoint and transfer function estimation from a rendered image.

\section*{Acknowledgement}
This work was supported by the National Key Research \& Development Program of China (2017YFB0202203), National Natural Science Foundation of China (61472354, 61672452 and 61890954), and NSFC-Guangdong Joint Fund (U1611263).

\bibliographystyle{ACM-Reference-Format}
\bibliography{template}

\end{document}